\documentclass[journal, onecolumn]{IEEEtran}
\IEEEoverridecommandlockouts
\renewcommand\IEEEkeywordsname{Keywords}
\usepackage{url}
\usepackage{amsmath}
\usepackage{amsfonts}
\usepackage{graphicx}
\usepackage{subfigure} 
\usepackage{xcolor}
\usepackage{multirow}
\usepackage{array}
\usepackage{ntheorem}
\usepackage{booktabs}
\usepackage{adjustbox}
\usepackage{hyperref}
\usepackage{bm}

\usepackage{mdframed}
\usepackage{framed}  
\usepackage{multicol}

\begin{document}

\title{Analyzing the Impact of Design Factors on Solar Module Thermomechanical Durability Using Interpretable Machine Learning Techniques
\thanks{The project was primarily funded and intellectually led as part of the Durable Modules Consortium (DuraMAT), an Energy Materials Network Consortium funded under Agreement 32509 by the U.S. Department of Energy (DOE), Office of Energy Efficiency \& Renewable Energy, Solar Energy Technologies Office (EERE, SETO). Lawrence Berkeley National Laboratory is funded by the DOE under award DE-AC02-05CH11231.
\newline
\newline
The authors declare no conflicts of interest. The views expressed in the article do not necessarily represent the views of the DOE or the U.S. government. Instruments and materials are identified in this paper to describe the experiments. In no case does such identification imply recommendation or endorsement by LBL. The U.S. government retains and the publisher, by accepting the article for publication, acknowledges that the U.S. government retains a nonexclusive, paid-up, irrevocable, worldwide license to publish or reproduce the published form of this work, or allow others to do so, for U.S. government purposes.}
}

\author{
\IEEEauthorblockN{Xin Chen~\IEEEauthorrefmark{1}\IEEEauthorrefmark{2}, Todd Karin~\IEEEauthorrefmark{3}, Anubhav Jain~\IEEEauthorrefmark{1} \vspace{15pt}} \\
\IEEEauthorblockA{
\IEEEauthorrefmark{1}Lawrence Berkeley National Laboratory, Berkeley, CA, U.S.A \\
\IEEEauthorrefmark{2}University of California, Berkeley, Berkeley, CA, U.S.A \\
\IEEEauthorrefmark{3}Kiwa PVEL, Member of Kiwa Group, Napa, CA, U.S.A
}}

\maketitle

\begin{abstract}
Solar modules in utility-scale systems are expected to maintain decades of lifetime to rival conventional energy sources. However, cyclic thermomechanical loading often degrades their long-term performance, highlighting the importance of effective design to mitigate thermal expansion mismatches between module materials. Given the complex composition of solar modules, isolating the impact of individual components on overall durability remains a challenging task. In this work, we analyze a comprehensive data set that comprises bill-of-materials (BOM) and thermal cycling power loss from 251 distinct module designs to identify the predominant design factors and their impacts on the thermomechanical durability of modules. The methodology of our analysis combines machine learning modeling (random forest) and Shapley additive explanation (SHAP) to correlate design factors with power loss and interpret the model's decision-making. The interpretation reveals that silicon type (monocrystalline or polycrystalline), encapsulant thickness, busbar numbers, and wafer thickness predominantly influence the degradation. With lower power loss of around 0.6\% on average in the SHAP analysis, monocrystalline cells present better durability than polycrystalline cells. This finding is further substantiated by statistical testing on our raw data set. The SHAP analysis also demonstrates that while thicker encapsulants lead to reduced power loss, further increasing their thickness over around 0.6 to 0.7mm does not yield additional benefits, particularly for the front side one. In addition, other important BOM features such as the number of busbars are analyzed. This study provides a blueprint for utilizing explainable machine learning techniques in a complex material system and can potentially guide future research on optimizing the design of solar modules. 
\end{abstract}


\begin{figure}[h]
    \centering
\begin{mdframed}
    \begin{multicols}{2}
        \textbf{Nomenclature}
        \begin{description}
        \labelsep1.5em
          \item[PV] Photovoltaic
          \item[BOM] Bill of materials
          \item[TC] Thermal cycling
          \item[CTE] Coefficient of thermal expansion
          \item[PQP] Product qualification program
          \item[Encaps] Encapsulant
          \item[Glass2] Rear glass of glass-glass modules
          \item[Mono-c] Monocrystalline
          \item[Poly-c] Polycrystalline
          \item[EVA] Ethylene vinyl acetate
          \item[POE] Polyolefin elastomer
          \item[TPT] Tedlar Polyester Tedlar
          \item[ML] Machine learning
          \item[KNN] K-nearest neighbors
          \item[SVR] Support vector regression
          \item[RF] Random forest
          \item[RMSE] Root mean square error
          \item[SHAP] SHapley Additive exPlanation
          \item[$\phi$] Shapley value
          \item[CI] Confidence interval
          \item[ANOVA] Analysis of variance
        \end{description}
    \end{multicols}
\end{mdframed}
\end{figure}

\section{Introduction} \label{introduction}
Utility-scale photovoltaic (PV) systems are expected to achieve an extended operating lifetime to be competitive with conventional energy sources~\cite{Ardani2022solar}. However, solar modules installed in the field are subject to multiple environmental stresses such as ultraviolet light, temperature variation, mechanical loading induced by snow, wind, hail~\cite{aghaei2022review}. These factors introduce multiple pathways of degradation, reducing the durability of the module. One of the sources of long-term degradation is the cyclic thermomechanical deformation of solar modules caused by temperature variation. Over time, thermal cycling can cause the degradation of components within solar modules, such as interconnections, and lead to a decrease in module power~\cite{aghaei2022review, jordan2017photovoltaic}. Therefore, it is essential to identify the potential problems in the current design of solar modules and optimize module robustness to thermal cycling degradation. 

\begin{figure}
\centering
\includegraphics[width=0.75\textwidth]{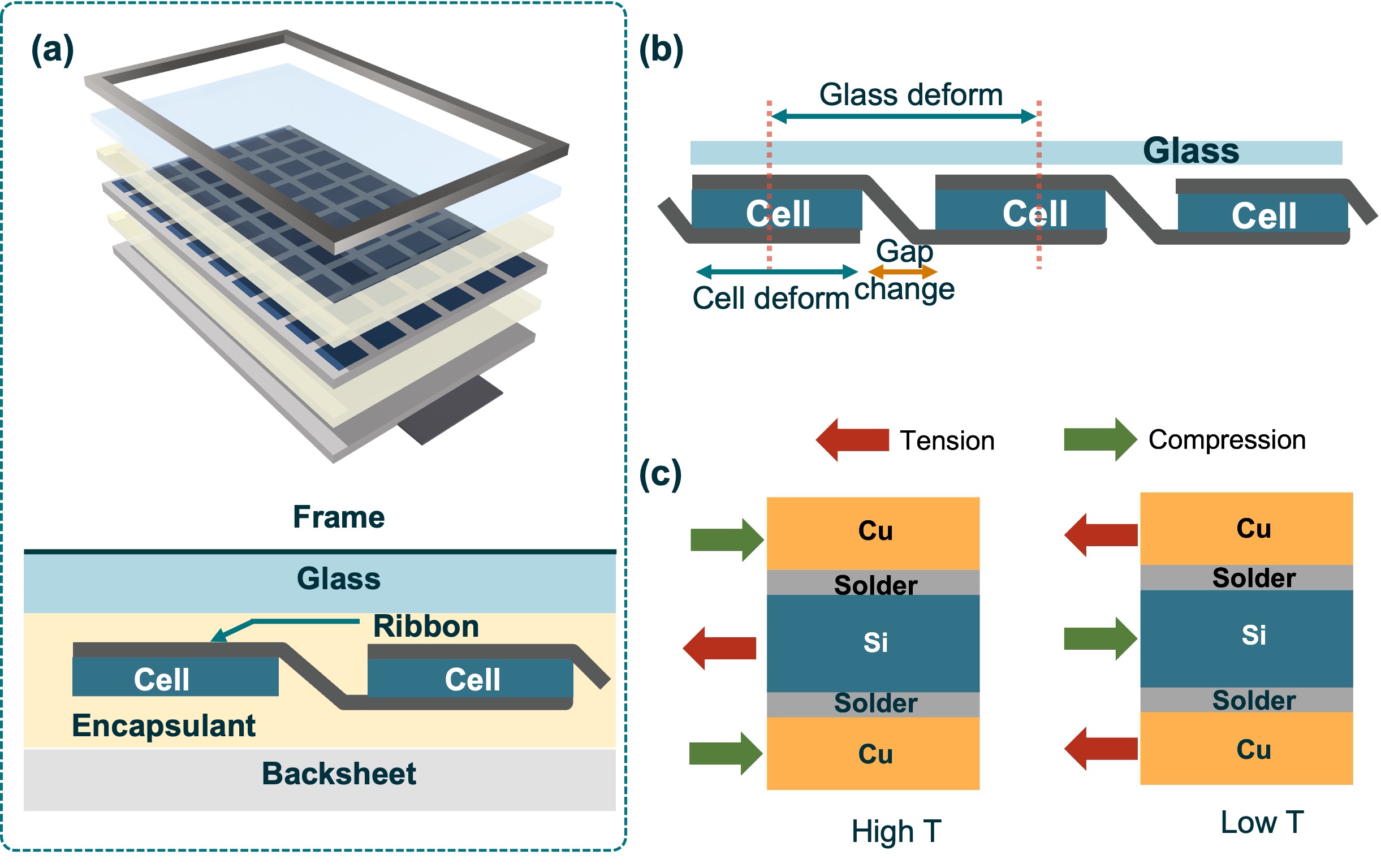}
\caption{Schematic structure of a glass/backsheet PV module and the mechanism of its thermomechanical degradation. (a) The layered structure of a PV module. (b) Mechanism of the cyclic deformation of ribbons. (c) Mechanism of cyclic thermal stress in solder layer.}
\label{fig:module}
\end{figure}

Figure~\ref{fig:module}(a) illustrates an example of a typical glass/backsheet module. Its multi-layered construction comprises a front frame, a glass layer, polymer encapsulant layers commonly made from ethylene vinyl acetate (EVA) or polyolefin elastomer (POE), a solar cell layer composed of silicon solar cells and copper interconnections, and a backside polymer backsheet made from layers of Tedlar Polyester Tedlar (TPT). The copper ribbons are connected to the cell metallization with solder (SnPb). A defining characteristic of these components is their distinct thermal expansion coefficients (CTE, $\alpha$). This variance in CTEs means that they expand or contract at different rates in response to daily and seasonal temperature fluctuations in their operating environment. 

Previous research~\cite{kawai2017causes} showed that the power loss of different models of PV modules after extended thermal cycling tests of up to 600 cycles varied from 0.8\% to 14.5\%. This degradation is partially due to the CTE mismatch between glass ($\alpha_\text{glass}\approx8\times10^{-6}\text{K}^{-1}$) and Si cells ($\alpha_\text{cell}\approx2.5\times10^{-6}\text{K}^{-1}$). The cells adhere to the glass by encapsulants, so the CTE mismatch causes a non-uniform in-plane displacement of cells and glass~\cite{eitner2010use}. The gap between cells can vary with different temperatures, as shown in Figure~\ref{fig:module}(b). Such temperature-driven variations over the lifetime of a module can result in cyclic deformation of the copper ribbons that connect cells and cause interconnection fatigue~\cite{eitner2010use,borri2018fatigue, wiese2010mechanical}. Additionally, fragments in cracked cells can shift due to temperature variation, leading to wear and tear at metal contacts~\cite{silverman2019movement}. Another prevalent degradation mode caused by temperature variation is solder disconnection. The CTE mismatch between the copper ($\alpha_\text{Cu}\approx16.7\times10^{-6}\text{K}^{-1}$) ribbon and Si cell ($\alpha_\text{Si}\approx2.5\times10^{-6}\text{K}^{-1}$) triggers non-uniform deformations in these layers~\cite{park2022reliability}. As depicted in Figure~\ref{fig:module}(c), at elevated temperatures, Si and Cu experience tensile and compressive stresses, and at low temperatures, the stresses are reversed, which causes a periodic change of shear stress within the solder layer~\cite{park2022reliability}. This cyclic loading in the solder often culminates in solder disconnection~\cite{jeong2012field, itoh2014solder, hanifi2020loss}. Solder degradation is one of the failure modes that occurs in the early stage of module operation, which can be probed by 600 cycles of thermal cycling accelerated aging tests~\cite{kawai2017causes}.

As shown above, thermal cycling degradation involves multiple modes, strongly impacted by the specific bill-of-materials (BOM) of solar modules. This encompasses the dimensions and material properties of each module component. Understanding which design factors predominantly affect thermal cycling power loss is of great significance in guiding future research on module optimization. Several previous studies have investigated the impact of various design factors. Bosco \emph{et al.}~\cite{bosco2016influence} performed a regression analysis to identify design factors that may influence the degradation of the solder by simulating the accumulated damage to the solder joints during thermal cycles. The top sensitive factors are the thickness of the solder layer, Cu ribbon, and Si wafer layer. Park \emph{et al.}~\cite{park2022reliability} also confirmed that reducing the thickness of the Si cell and copper ribbon can increase the lifetime of the solder in thermal cycling by simulation. Zhu \emph{et al.}~\cite{zhu2018effect} fabricated different mini-modules and did a simulation to investigate the effect of viscoelasticity of encapsulant materials on the solder joint fatigue. They found that modules with encapsulants of higher viscous properties presented more power loss after the thermal cycling test. Beinert \emph{et al.}~\cite{beinert2019effect} found that increasing cell size and changing full cells into half-cut cells could decrease thermal stress on the cell layer using simulation. They further qualitatively concluded that cell thickness, encapsulant CTE and glass/backsheet CTE strongly influence the stress in the cell layer~\cite{beinert2022thermomechanical}. Hanifi \emph{et al.}~\cite{hanifi2020loss} also found that the higher rigidity of encapsulant materials and changing full cells into half-cut cells could mitigate ribbon fatigue during temperature variation. As exemplified by this past research, most sensitivity analysis work in this field relies on simulated data. Computational constraints often force researchers to model simplified structures, such as a single cell rather than a full-size solar module, or ignore the busbars, which may bypass some real-world effects. 

Previous research~\cite{bosco2016influence} correlated module design with durability using a linear model to explore the effects of these design factors. Typically, the standard regression analysis with linear model imposes prerequisites on the data set, such as multivariate normality, homoskedasticity in the data. In addition, the linear model requires the linear relationship between the independent variables and the target variable; otherwise transformations of the independent variables such as logarithm and reciprocal transformation~\cite{bosco2016influence} are necessary to capture the non-linear relationship. On the other hand, machine learning (ML) models are mostly nonparamatric and present an outstanding ability to discover the underlying pattern in the data. While the enhanced predictive accuracy of advanced ML models is commendable compared to linear models, a significant challenge emerges in their interpretability. Several methodologies have been proposed to solve this trade-off between accuracy and interpretability. Noteworthy among these are Partial Dependence Plot (PDP)~\cite{friedman2001greedy}, Local Interpretable Model-agnostic Explanations (LIME)~\cite{ribeiro2016should}, and SHapley Additive exPlanations (SHAP)~\cite{lundberg2017unified}. In particular, SHAP is able to do both global and local interpretation, and its adoption across diverse scientific disciplines is a testament to its efficacy~\cite{bannigan2023machine, giles2022machine, kumar2021chemical, hartono2020machine}. These model-agnostic methodologies facilitate data inference with more predictive machine learning models beyond traditional linear models, whilst preserving interpretability.

This study identifies the predominant BOM features that affect the thermomechanical durability of modules and explores their impacts using real-world data. Our data set comprises full-size module data from the industry that includes both BOM features and the corresponding power loss after 600 standard thermal cycles~\cite{international2016photovoltaic}. Since these are existing records from the industry rather than from a well-designed data sampling process, we propose using nonparametric machine learning models to correlate the BOM features with the power loss and then applying the model-agnostic method to interpret the black-box models. By comparing various ML models, we develop a random forest (RF) model~\cite{ho1995random} with a testing root mean square error (RMSE) of 1.179\%. Subsequently, we employ the SHAP method to interpret the developed model, thus identifying the predominant BOM features as Si type, encapsulant thickness, number of busbars, and wafer thickness. Also, with SHAP analysis, we elucidate the impacts of these top important factors on the thermomechanical durability. In the end, we apply statistical testing on the original data set to verify our conclusions. 

\begin{figure*}
\centering
\includegraphics[width=1\textwidth]{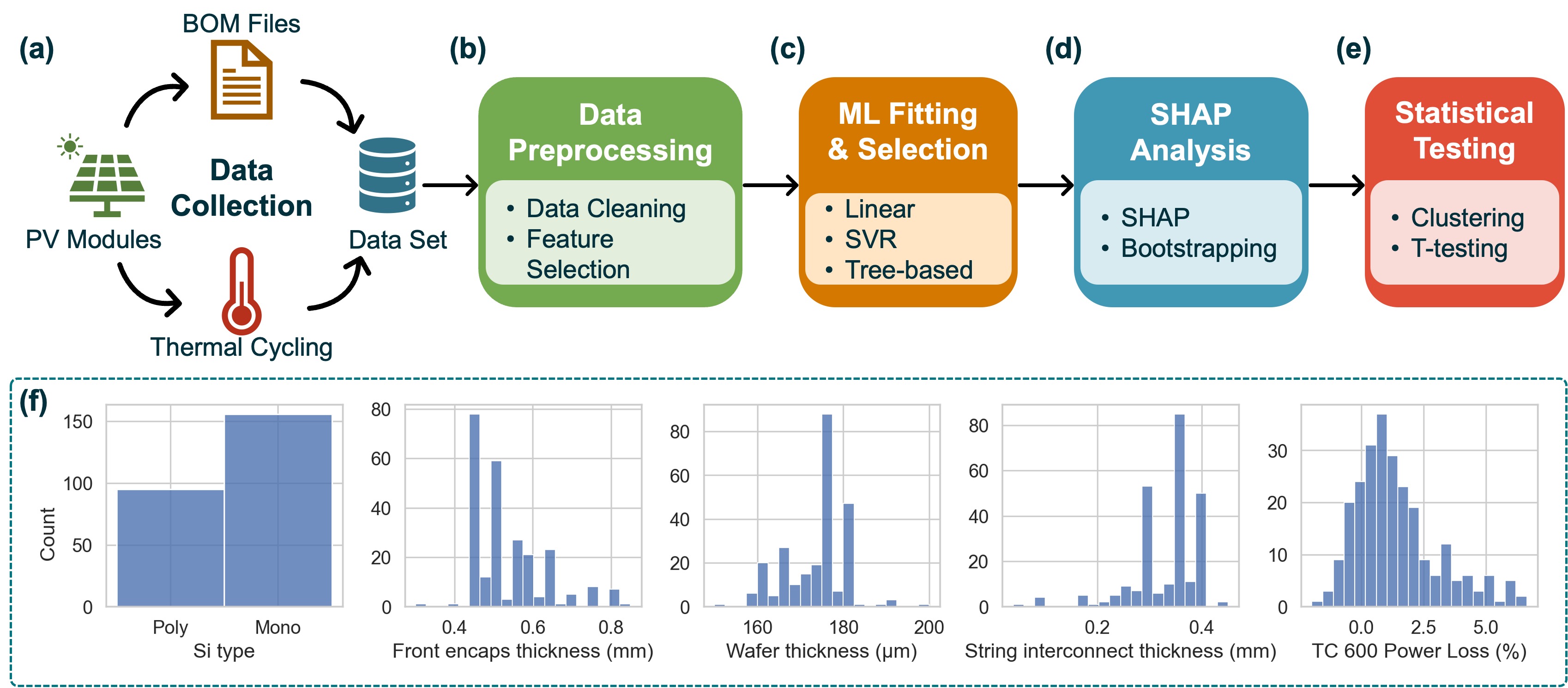}
\caption{The workflow of the study and data distribution. (a)-(e) illustrate the steps performed in this study. (a) Data collection. The data set is constructed by manually extracting information from BOM files and measuring the power loss under thermal cycling tests. (b) Data preprocessing. The data set is then cleaned, and the corresponding features are selected to build machine learning models. (c) ML model selection. Multiple models such as linear models, support vector regressor (SVR), and tree-based models are fitted and compared to correlate the BOM features with the TC 600 power loss. The best-performing model is selected for interpretation. (d) SHAP analysis. SHAP method is then used to interpret this optimal model to determine the predominant design factors and their impacts on the power loss. The uncertainty of Shapley values is quantified using bootstrapping. (e) Validation of the interpretation. In the end, post hoc statistical testing is used to validate the interpretation. Clustering method is used to regroup the data set and t-testing is used to test the statistical significance. (f) Distribution of the selected feature values and the power loss of each module. The distribution of other features is shown in the SI.}
\label{fig:workflow}
\end{figure*}

To sum up, the schematic workflow of this project is demonstrated in Figure~\ref{fig:workflow}(a)-(e). The principal contributions of this study are summarized as follows:
\begin{enumerate}
\item This work identifies that Si type, encapsulant thickness, busbar numbers and wafer thickness are the predominant BOM features that influence module’s thermomechanical durability. 
\item We find that using mono-c Si cells presents lower TC power loss (0.6\%) than poly-c Si ones, and thicker encapsulant over 0.6-0.7mm does not bring additional benefits to the durability.
\item We propose a novel workflow to the field of understanding the complex correlations between BOM data and module durability. The methodologies combine machine learning modeling (RF model), model-agnostic interpretation method (SHAP), and post hoc validation with statistical testing.
\item We publish the processed BOM data and codes in our work. This data set can be beneficial for future research on improving the durability of PV modules considering the difficulty in tracking the BOM features. Some values in this public data set are redacted due to non-disclosure agreement.

\end{enumerate}

The rest of this article is organized as follows: Section~\ref{methods} illustrates the details of thermal cycling setup and computation methodologies, including data collection, the feature matrix constructed by feature selection, ML modeling, SHAP interpretation and statistical testing. Section~\ref{results} reports the results derived from the application of these methods. We first compare the performance of ML models and report the generalization of the optimal RF model. After that we rank the importance of each BOM features and demonstrate the impact of predominant features on the durability by interpreting the aforementioned RF model with SHAP analysis. We further discuss the underlying mechanisms of the connections between BOM and the module durability. In the end, we validate the interpretation of the ML model using statistical testing. Section~\ref{conclusion} summarizes this paper and provides further insight into the current module design and the durability of modules. 

\section{Methodology} \label{methods}
All experimental data were collected at PVEL~\cite{pvel}. All codes and models compiled in this research were written in Python3~\cite{python3} and executed on a MacBook Pro (Apple M2 Pro chip, 16 GB memory).

\subsection{BOM Data Collection}
The data comes from PVEL's product qualification program (PQP)~\cite{PVELcard}. 275 full-size commercial solar modules of different designs were collected from 47 different manufacturers by PVEL. These modules are estimated to cover around 70\% various types of designs in the market by the date of collection. The BOM data were extracted manually from the document files provided by these manufacturers and yielded over 100 features, including supplier information, dimensions and configurations, materials of PV components such as cells, connections, encapsulants, backsheets and glasses. In this study, we select certain features that are essential for machine learning modeling based on the steps described in the following Section~\ref{data-preprocessing}.

\subsection{Thermal Cycling Test}
The thermal cycling test is carried out at PVEL according to the standard outlined in IEC 61215-2:2016~\cite{international2016photovoltaic}. Each module is placed in an environmental chamber and subjected to temperature cycles from $-40^\circ\text{C}$ to $+85^\circ\text{C}$. There are 600 thermal cycles, and each cycle contains a maximum $100^\circ\text{C}/\text{hr}$ ramp rate and a minimum dwell time of 10 minutes. Each complete cycle takes around 6 hours. A current-voltage (IV) flash test (Pasan SunSim 3B) is performed under standard test conditions following IEC 60904-1:2006~\cite{international2006photovoltaic} before the aging test and after every 200 cycles to measure the IV curve of the modules.

\subsection{Data Preprocessing and Feature Selection} \label{data-preprocessing}
The final feature matrix $\mathbf{X}$ we constructed to train the machine learning models contains 251 modules with 22 BOM feature columns and the target variable $\mathbf{Y}$ is the power loss (\%) after 600 thermal cycles. The distribution of some selected BOM features, along with the power loss after 600 cycles of thermal cycling test (TC 600), is plotted in Figure~\ref{fig:workflow}(b). The complete distribution of the 22 features can be found in the SI Section ``Feature Selection''. The power loss is defined in Equation~\ref{powerloss}. 
\begin{equation}
\Delta P(\%) = \frac{P^{max}_{init} - P^{max}_{after}}{P^{max}_{init}} \times 100\%
\label{powerloss}
\end{equation}
where $P_{init}^{max}$ is the module's maximum output power before the aging test and $P_{after}^{max}$ is the maximum output power after 600 thermal cycles.

To construct this feature matrix (\emph{i.e.,} the input X), we preprocess the raw data by data cleaning and feature selection. Data cleaning includes outlier detection using residual analysis, data type casting, and missing data handling using  a K-nearest neighbors (KNN) imputer~\cite{troyanskaya2001missing}. The outlier detection step removes 24 modules that do not have TC 600 power loss data or have large residuals (over 97\% quantile) in the residual analysis, as demonstrated in the Supporting Information (SI) section titled ``Data Preprocessing''. Detailed descriptions of these preprocessing steps are also included in this SI section.

\begin{figure*}
\centering
\includegraphics[width=1\textwidth]{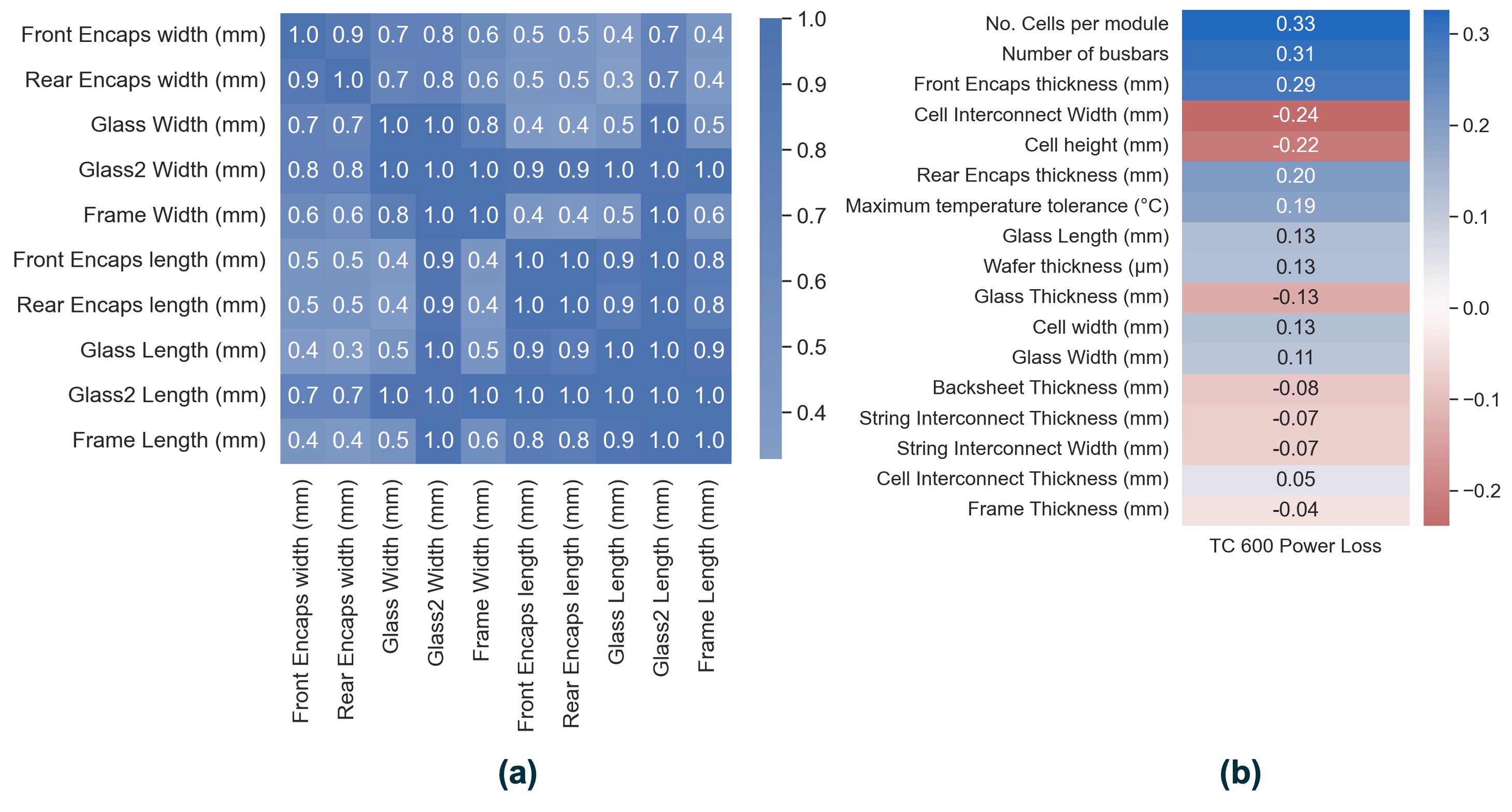}
\caption{The Spearman correlation between numerical features and the target variable is calculated to assist the construction of the feature matrix in combination with domain knowledge. The correlations shown here are solely used as a reference for feature selection and do not necessarily reflect the true correlations or importance since multiple variables may be entangled. (a) Correlation matrix of numerical features related to module dimensions. (b) Spearman correlation score between selected features and power loss. Features with high correlation coefficient such as the number of busbars are used to construct the feature matrix, although they were not explored in some previous analyses~\cite{bosco2016influence, beinert2022thermomechanical}. The full correlation matrices of numerical features as well as categorical features are shown in the SI.}
\label{fig:correlation}
\end{figure*}

Subsequent to the data cleaning process, we select features from more than 100 BOM attributes in the raw data set. The goal of feature selection is to identify features with potential correlations to the target variable among other BOM features in the original data set and reduce inter-feature dependencies because high correlation among features makes the interpretation of feature importance less accurate and less stable. This selection is primarily guided by domain-specific knowledge. According to previous research~\cite{bosco2016influence, beinert2019effect, beinert2022thermomechanical}, the features that influence the stress distribution in solar modules include the thickness, width, and length of each layer and the interconnection thickness, so these features are included in the feature matrix. Mechanical properties such as viscoelasticity of the polymer encapsulant may also influence the durability of the module, but these properties are difficult to track in the BOM data and are not discussed in this work. We also perform exploratory data analysis using a correlation matrix to further test the association among features and between features and power loss. These associations are calculated using the Python package Pingouin~\cite{vallat2018pingouin} and Scipy~\cite{2020SciPy-NMeth}. 

We employ different statistics to measure the associations for different types of variables. We use Spearman correlation to test the correlation between two numerical variables $X$ and $Y$. For instance, Figure~\ref{fig:correlation}(a) is a correlation matrix that shows some highly dependent features related to module dimensions, so these features apart from glass length and width are excluded from the feature matrix to mitigate multicollinearity. We explain the impact of dependent features on feature importance in the SI section ``Feature Selection''. Furthermore, we select features that may not be explored in previous regression analyses~\cite{park2022reliability, bosco2016influence, beinert2022thermomechanical} but still show correlation with thermal cycling degradation based on the correlation matrix in Figure~\ref{fig:correlation}(b). For example, the number of busbars was not considered in the previous simulation~\cite{bosco2016influence} to save computation time, but we still include this feature in our models because this feature shows a high Spearman correlation in this data exploration process. It should be noted that the Spearman correlation is applicable to numerical variables. Regarding other types of variables, we use the Chi-square test~\cite{bruce2020practical} to test the statistical association among categorical features, and Analysis of Variance (ANOVA)~\cite{bruce2020practical} to test the statistical association between numerical features and categorical features as detailed in the SI Section ``Feature Selection''. We note that the dependence computed here is only used to help construct the feature matrix and does not necessarily reflect the true correlation between features and power loss due to the interaction of features. Therefore, further analysis like statistical analysis or machine learning modeling is required to determine correlations or feature importance. 

\subsection{ML Models}
In this study we compare multiple regression models, including linear (Lasso and ridge regression), support vector regression, and tree-based (RF and XGBoost) models to find the model that minimizes prediction error. Linear models act as a baseline for model comparison. We use k-fold ($k=5$) cross-validation to test the robustness. The hyperparameters of these models are tuned by using grid search, as described in the SI. In all cases, the model performance is assessed using root mean square error (RMSE) between the measured values and the predicted values. We build ML models using toolkits from the Scikit-learn package~\cite{scikit-learn} to model the correlation between BOM features and the power loss. The XGBoost model is not originally included in Scikit-learn and is constructed using the open source XGBoost package~\cite{chen2016xgboost}. The split of the development set, the model evaluation metric, and the description of model architectures are illustrated as follows:

\subsubsection{Data Splitting and Model Evaluation}
The data set is split into the training set and the testing set with a splitting ratio of 8:2. The training set is further divided into five parts for 5-fold cross-validation to compare models and do hyperparameter tuning. In each validation iteration, four parts are used for training and the rest part is used for validation. The performance after cross-validation is the average performance from each iteration. Furthermore, the hold-out testing set is used to evaluate the generalization of the optimal model picked from the cross-validation process. The root mean squared error (RMSE) used to evaluate model performance is defined in Equation~\ref{rmse}.
\begin{equation}
RMSE=\sqrt{\frac{1}{n}\sum_{i=1}^n {\left(\hat{y}_i-y_i\right)^2}}
\label{rmse}
\end{equation}
where $n$ is the number of data points, $\hat{y}_i$ is the prediction for the data point $i$ and $y_i$ is the ground truth of $i$. The RMSE shares the same unit (\%) as the power loss.

\subsubsection{Linear Model}
The generalized formula of a linear model follows Equation~\ref{linear}:
\begin{equation}
\mathbf{y} = \mathbf{X w}
\label{linear}
\end{equation}
where $\mathbf{y} \in \mathbb{R}^n$ is the target vector, $\mathbf{X} \in \mathbb{R}^{n \times d}$ is the feature matrix, and $\mathbf{w} \in \mathbb{R}^d$ is the model weight. $n$ is the number of measurements and $d$ is the number of features. The training aims to find the $\mathbf{w}$ that minimizes the loss between the measured values and the predicted values, as shown in Equation~\ref{argmin}.
\begin{equation}\hat{\mathbf{w}}=\operatorname{argmin}_{\mathbf{w}}\|\mathbf{y}-\mathbf{X w}\|_2^2+\lambda R
\label{argmin}
\end{equation}
where $\lambda R$ is the regularization term with the hyperparameter $\lambda$ to prevent overfitting. Lasso regression~\cite{tibshirani1996regression} with $L1$ regularization ($\lambda\|\mathbf{w}\|_1$) and Ridge regression~\cite{hoerl1970ridge} with $L2$ regularization ($\lambda\|\mathbf{w}\|_2^2$) were trained. The equations for these regularization terms are shown in the SI Section ``ML Modeling''.

\subsubsection{SVR}
Support vector regression~\cite{drucker1996support} finds a hyperplane with margins that minimize the error between the true value and predicted values. The equations for SVR is shown in the SI Section ``ML Modeling''.

\subsubsection{Tree-based Model}
Random forest and XGBoost are trained in this research. Random forest regression builds an ensemble of decision trees on different subsets of the original data using bootstrapping and splits nodes on a random subset of features, helping to increase robustness and prevent overfitting. The final prediction result is the average of the predictions of each decision tree. XGBoost uses boosting method by building decision trees sequentially, each trained to correct its predecessor's errors. Both tree-based models can be regularized by limiting the depth of trees or the number of nodes. XGBoost can also be regularized using $L1$ or $L2$ regularization in gradient boosting.

\subsection{SHAP Analysis} \label{method: shap}
To interpret ML models and understand the correlation between BOM features and power loss, we use SHAP method to explain the ML models and apply bootstrapping to quantify the uncertainty of this method. SHAP analysis reveals the impact of a certain feature on model prediction while marginalizing the effect of other features. A prediction (\emph{i.e.,} power loss) of the model can be considered as the sum of the contributions (\emph{i.e.,} Shapley values) of each BOM feature. The average contributions of each BOM feature represent the feature importance, and the dependence between feature values and Shapley values reflects the feature impacts on the power loss. 
\subsubsection{SHapley Additive exPlanations}
SHAP is a model-agnostic method that interprets each feature's marginal contribution to a specific prediction of the machine learning model. It can interpret both local prediction and global contributions of each feature via the additive method. SHAP measures the marginal contribution by computing the Shapley value of each feature ($\phi_j$), defined as:
\begin{equation}
\phi_j=\sum_{S \subseteq\{1, \ldots, p\} \backslash\{j\}} \frac{|S| !(p-|S|-1) !}{p !}(\text{Val}(S \cup\{j\})-\text{Val}(S))
\label{shap}
\end{equation}
where $S$ is a subset of the features used in the model, and $p$ is the number of features. $\{1, \ldots, p\} \backslash\{j\}$ represents the set without feature $j$. $\text{Val}(S)$ is the prediction for feature values in set $S$ that are marginalized over features not included in set $S$.

A prediction can be interpreted as the sum of Shapley values of each feature, as shown in Equation~\ref{linear_shap}, where $\hat{f}(\bm{x})$ is the prediction, $E_X[\hat{f}(X)]$ is the expectation of the prediction,  $\phi_j$ is the Shapley value (\emph{i.e.,} contribution) of the feature $j$, $\bm{x}$ is a feature vector that contains feature values in this prediction. Equation~\ref{linear_shap} illustrates that each Shapley value is correlated with the prediction, so the impact of a feature on its Shapley value also reflects its impact on the target variable.
\begin{equation}
\hat{f}(\bm{x})=E_X[\hat{f}(X)]+\sum_{j=1}^p \phi_j
\label{linear_shap}
\end{equation}

\subsubsection{Bootstrapping}
The uncertainty of the Shapley value is determined by using Bootstrapping~\cite{efron1992bootstrap}, which randomly sampled with replacement from the original data set 1,000 times and Shapley values of each feature are computed in each iteration. This results in the distribution of the Shapley value of each feature. The confidence interval (CI) of the Shapley value is quantified by using a 95\% confidence level.

\subsection{Post hoc Statistical Analysis}
Utilizing SHAP analysis allows us to understand the most influential BOM features and their impacts on power loss. However, the validity of SHAP interpretation is contingent upon the performance of the machine learning models, necessitating supplementary investigations. The straightforward method of validation is to conduct an independent post hoc statistical testing on the power loss in the original data set. For instance, we can directly compare the power loss of poly-c Si modules and mono-c Si modules to test which Si type is more durable. However, since our data set comprises modules from various manufacturers, other BOM features are not controlled when we compare different Si types. Therefore, we need to reorganize the original data set. The steps include regrouping the original data set to generate control groups using the clustering method and subsequent statistical testing.
\subsubsection{Clustering}
K-means~\cite{kmeans} implemented with Scikit-learn is used to do the clustering, which partitions the data into a predefined number of clusters ($K$) by iteratively assigning each data point to the nearest centroid,  and then recalculating the centroids as the mean of all the points in the cluster until the centroids stabilize or a maximum number of iterations is reached. The ``elbow'' method is used to determine the optimal number of clusters to be used. Before clustering, the feature matrix is standardized following Equation~\ref{standardize} and weights are assigned to each feature based on their importance following Equation~\ref{weights}
\begin{equation}
    Z = \frac{X - \mu}{\sigma}
    \label{standardize}
\end{equation}
where $Z$ is the standardized feature matrix, $X$ is the training feature matrix, $\mu$ and $\sigma$ are the training samples' mean and standard deviation.

\begin{equation}
\text{feature}_i^* = \text{feature}_i \times \frac{\text{SHAP}_i}{\sum_j{\text{SHAP}_j}} 
\label{weights}
\end{equation}
where $\text{feature}_i$ is the standardized value of feature $i$ and $\text{SHAP}_i$ is the mean absolute Shapley value of feature $i$.

\subsubsection{T-test}
T-test is conducted using the Pingouin package~\cite{vallat2018pingouin}. The t-value for the group $a$ and $b$ is defined in the SI Section ``Post hoc Statistical Analysis''.

\section{Results and Discussion} \label{results}

\subsection{ML Modeling} \label{res:ml}
\begin{figure}
\centering
\includegraphics[width=0.95\textwidth]{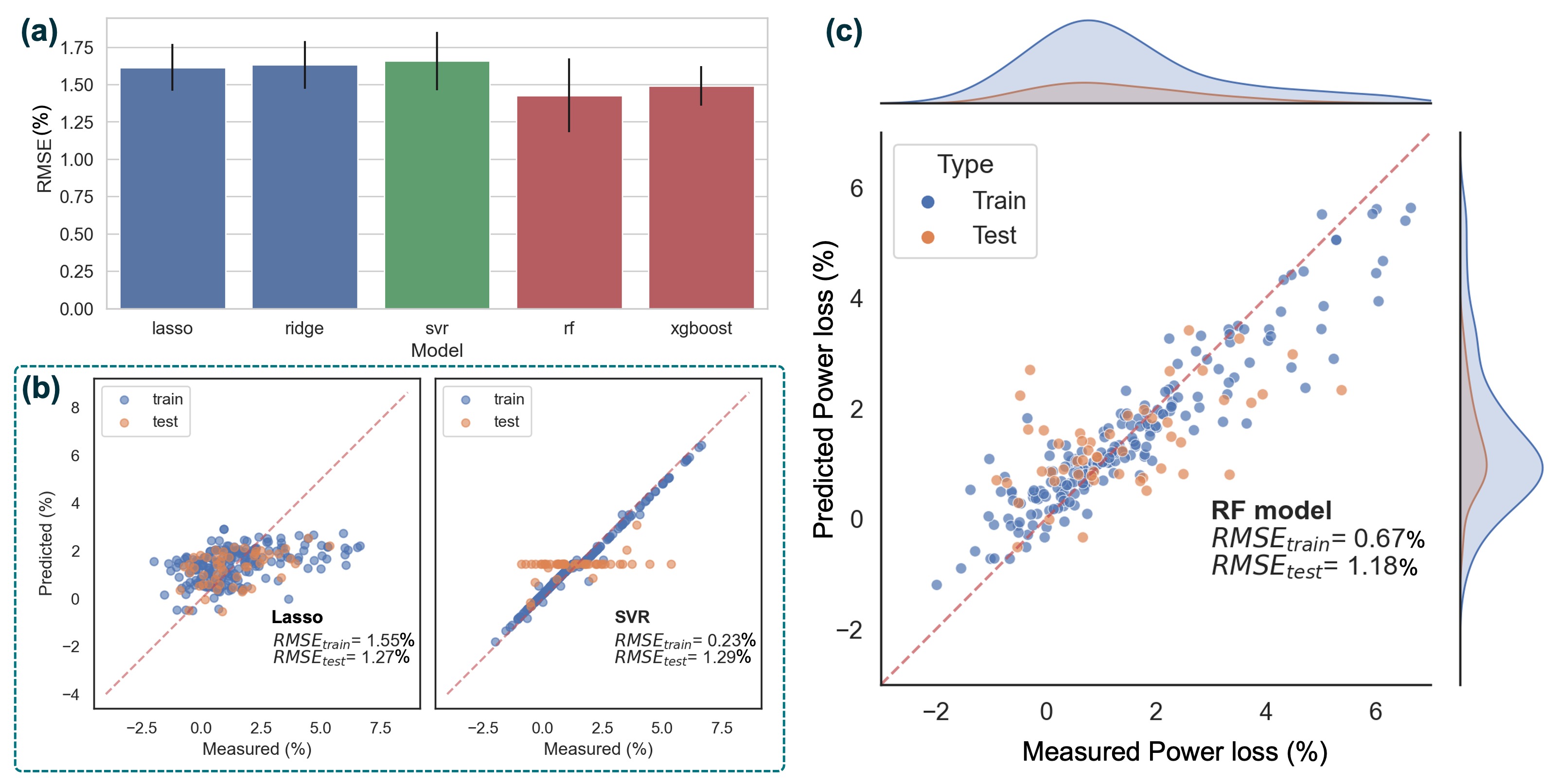}
\caption{Comparison of machine learning models and generalization test of the optimal model. (a) The 5-fold cross-validation RMSE score for each model. The error bar shows the standard deviation of scores in each fold. Tree-based models (RF and XGBoost) present the best performance. (b) The performance of the Lasso and SVR on the training and testing set, which present underfitting and overfitting. (c) Performance of the selected RF model in the hold-out testing set. Relatively low testing RMSE compared to the training score shows that the RF model is not severely overfitted and generalized well on the new data set.}
\label{fig:ml}
\end{figure}

In this work, we train a random forest model by taking the constructed BOM feature matrix as input and the TC 600 power loss as target variable. This model shows a validation RMSE of 1.427\% which outperforms other models in our study and a testing RMSE of 1.179\%, which illustrates the generalization of this model. It should be noted that the unit (\%) of RMSE is the same as the power loss. Before selecting this model, we compare several ML models as illustrated in Section~\ref{methods}. Table~\ref{tab:ml} and Figure~\ref{fig:ml}(a) compare the mean value and standard deviation of RMSE of each model during the 5-fold cross-validation. In general, tree-based models outperform other models, with RMSE validation of $1.427\% (\pm 0.247\%)$ for RF and $1.491\%(\pm 0.133\%)$ for XGBoost. This is unsurprising because linear models do not capture the nonlinear relationships between the BOM features and the power loss, and SVR tends to develop severe overfitting as shown in Figure~\ref{fig:ml}(b). Considering the low mean error and simpler implementation, we select RF as the optimal model for subsequent SHAP analysis. It should be noted that although the difference in RMSE values between tree-based models and other models may seem marginal in Table~\ref{tab:ml}, given the low target variable values, even minor RMSE variations can signify substantial prediction performance disparities. This significant difference in performance among models can be visualized in the scatter plots of predicted values \emph{vs.} measured values. For example, the plot for Lasso model in Figure~\ref{fig:ml}(b) clearly shows underfitting since the scatter points are not distributing along the ``Measured=Predicted'' line (the diagonal dash line). The scatter plots for all the models are demonstrated in the SI Section ``ML Modeling''. In addition, to mitigate the influence of large differences in ranges of features in the linear model, especially when these features have different units, we also train linear models on a standardized data set following the standardization equation (Equation~\ref{standardize}). However, the performance of the linear model is not varied (details included in SI). We note that tree-based models are not influenced by this difference between feature ranges and do not require standardized data set. 

\begin{table}
\centering
\caption{Mean value and standard deviation of RMSE of each model during the 5-fold cross validation.}
\begin{tabular}{|l|l|l|}
\hline
Model   & RMSE\_mean (\%) & RMSE\_std (\%) \\ \hline
Lasso (Linear)   & 1.614    & 0.156   \\ \hline
Ridge (Linear)  & 1.631    & 0.159   \\ \hline
SVR     & 1.657    & 0.195   \\ \hline
RF (Tree-based)     & 1.427    & 0.247   \\ \hline
XGBoost (Tree-based) & 1.491    & 0.133   \\ \hline
\end{tabular}
\label{tab:ml}
\end{table}

After comparing the RF model with other models, we further test its generalization on the hold-out 20\% of the data set (\emph{i.e.,} testing set), as shown in Figure~\ref{fig:ml}(c). The RMSE (1.179\%) of the hold-out testing data is not drastically higher than the training error, as opposed to the overfitted SVR model. This indicates that the fitted RF model is able to capture the underlying correlation between the BOM data and the degradation and has generalization to new data. Following model selection, we retrain the optimal RF model on the combined data set of both training and testing set, achieving a RMSE of 0.558\%, and proceed to interpret this model using the SHAP method.

\subsection{SHAP Analysis for Ranking BOM Feature Importance}
\begin{figure*}
\centering
\includegraphics[width=0.9\textwidth]{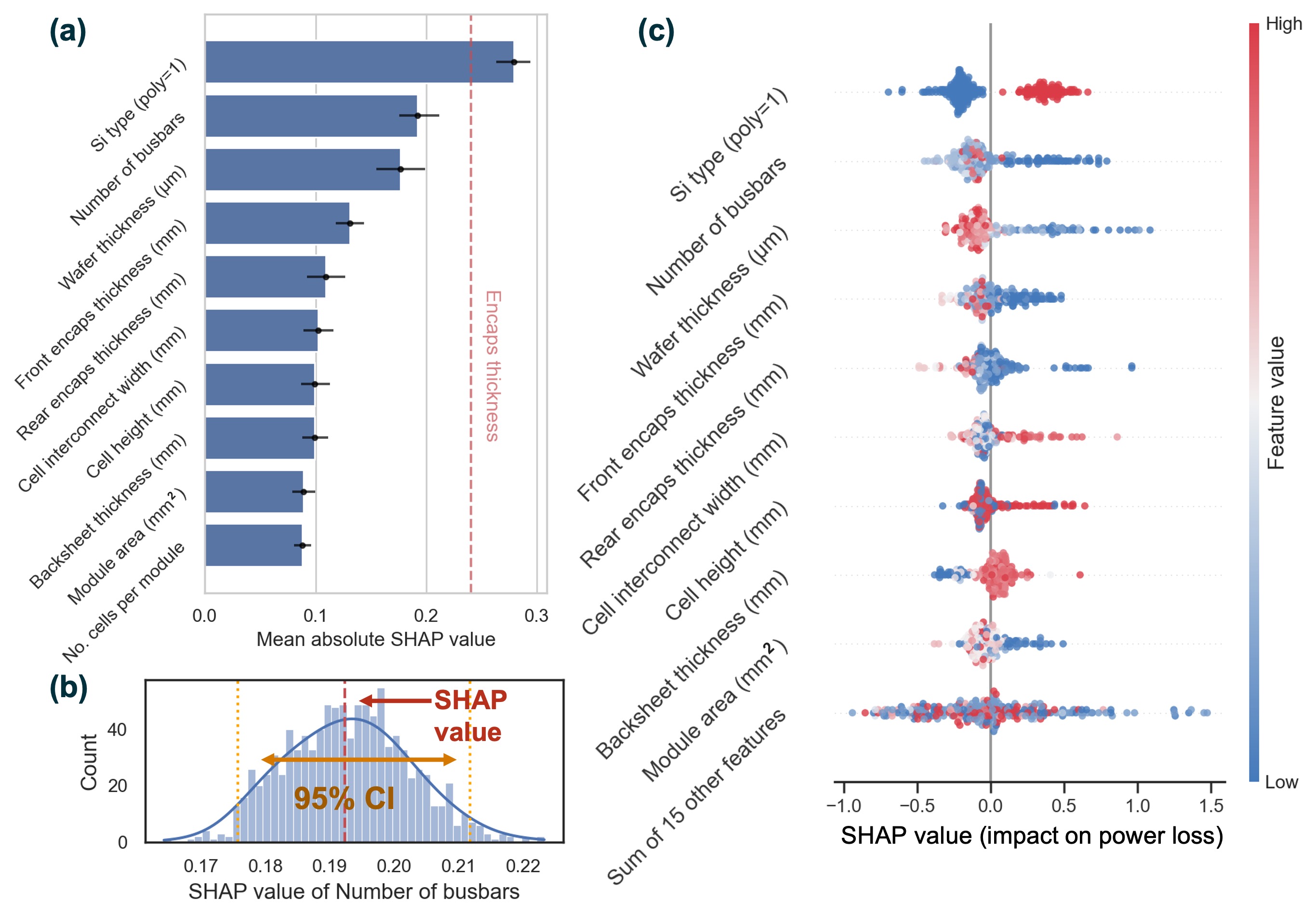}
\caption{SHAP interpretation of the machine learning model and uncertainty determination. (a) The rank of importance of the BOM features. The mean absolute Shapley value is calculated over all the modules in the original data set. The error bar shows the uncertainty with 95\% confidence level using bootstrapping (random sampling from the original data set). The red dashed line shows the sum of Shapley values of both front and rear encapsulant thickness, indicating that the encapsulant thickness is also a very important feature other than the Si type. (b) An example of the bootstrapping distribution of the number of busbars using 1000 sampling. The Shapley value computed from the original data is indicated by the red dashed line, and the boundaries of the 95\% CI computed from the bootstrapping are indicated by the orange dashed line. (c) The beeswarm plot of Shapley values. It shows the dependent relation between each feature and the power loss. The color of each point denotes the feature values. The red color denotes a larger feature value, and the blue color denotes a smaller feature. The x-axis is the Shapley value (\emph{i.e.}, impact on degradation) of each feature. For example, the blue dots (lower feature value) in the row ``Wafer thickness'' with positive Shapley values (higher power loss) suggest that a lower wafer thickness tends to increase power loss. }
\label{fig:shap}
\end{figure*}

To investigate the importance of features related to power loss, we use the SHAP method to interpret the RF model demonstrated in Section~\ref{res:ml}. With this method, we find that BOM features such as Si type, encapsulant thickness, busbar numbers and wafer thickness remain the top important factors, as shown in Figure~\ref{fig:shap}(a) which ranks the average impact of each feature on the prediction in the order of decreasing mean absolute Shapley values. In general, Figure~\ref{fig:shap} demonstrates that ``Si type (poly=1)'', which represents whether the solar cell is mono-c (denoted as 0) or poly-c (denoted as 1) Si, is listed as the most important feature. Also, this feature still remains the most important even when the lower bound of the confidence interval (shown as the error bars) is considered. This indicates that the Si type is the predominant design factor that impacts thermal cycling power loss. The subsequent influential features are ``Number of busbars'', ``Wafer thickness'', ``Front  encaps thickness'' and ``Rear encaps thickness''. If we consider the thickness of the front and rear encapsulant together due to their similarity and add up their Shapley values, the encapsulant thickness is the next important BOM feature other than the Si type as shown in the red dashed line in Figure~\ref{fig:shap}(a). 

To quantify the uncertainty of Shapley values, the 95\% confidence intervals are shown as the error bars in Figure~\ref{fig:shap}(a). The narrow interval indicates the robustness of this method to random selection in the training data. It should be noted that Si type and encapsulant thickness are the two top important features with robustness, but the importance rank of other features can vary according to the confidence interval in Figure~\ref{fig:shap}(a). Figure~\ref{fig:shap}(b) shows an example of the bootstrapping distribution of the Shapley value of the ``Number of busbars'', which follows a normal distribution consistent with the central limit theorem~\cite{bruce2020practical}.

\subsection{SHAP Analysis for Interpreting Impacts of BOM Features on Power Loss}
To understand the impact of varying a specific BOM feature on the power loss, we also examine the Shapley value of each feature in each measurement, as shown in Figure~\ref{fig:shap}(c). We will separately discuss the detailed interpretation of the impacts of each top important feature and possible mechanisms in the next few paragraphs. Due to the additive property of the SHAP formula shown in Section~\ref{methods}, the power loss of a module can be considered as the sum of Shapley values of each feature, and thus the dependence between the feature value and the Shapley value reflects the relation between the BOM feature and the power loss. For instance, the ``Si type'' row in Figure~\ref{fig:shap}(c) illustrates that poly-c Si modules (red clusters) possess positive Shapley values, which implies a tendency towards higher power loss for such modules.

The dependence between BOM features and power loss is clear for categorical variables such as ``Si type'' in Figure~\ref{fig:shap}(c), but it becomes challenging to understand this dependence for other numerical features in this plot. To better visualize this dependence between BOM features and power loss for numerical features, we also construct Figure~\ref{fig:dependence} based on Figure~\ref{fig:shap}(c) to illustrate the relationship between the feature values and their corresponding Shapley values.  The Shapley values ($<0.1$) of other features are not significant compared to these primary ones and are not discussed in detail in this paper. Figure~\ref{fig:dependence}(a)-(d) show the dependence plot of the top five important features besides the Si type and Figure~\ref{fig:dependence}(e)(f) demonstrate two features that were investigated in previous studies~\cite{park2022reliability, bosco2016influence}.  

\subsubsection{Impact of Si type}
Based on the SHAP interpretation, Si type is the predominant design factor regarding the module's thermomechanical durability and poly-c Si modules present higher power loss. Figure~\ref{fig:shap}(a) illustrates that using poly-c Si instead of mono-c Si increases the Shapley value by around 0.6 on average, which indicates an increase of 0.6\% in the power loss. We note that this value only quantifies the impact of using poly-c Si, but because of the noises in our data set and model, the change in power loss after replacing mono-c Si with poly-c Si may vary in reality. To the best of our knowledge, this difference in durability between mono-c and poly-c Si modules was not reported previously. Since mono-c and poly-c Si have similar CTE~\cite{watanabe2004linear}, the variance of power loss may not be caused by the difference in thermal expansion. One plausible reason involves the grain boundary in poly-c Si that serves as the initiation sites of micro-cracks~\cite{zavattieri2001grain}. These micro-cracks can propagate with successive thermal cycles, leading to a degradation in the performance of the solar module over time even before the crack length reaches the threshold of a catastrophic breakage. In addition, different cutting methods for processing the silicon ingot can influence the fracture toughness of Si wafers~\cite{wu2013effect}, so this manufacturing step can be a potential confounding variable that causes this phenomenon. We note that the impact of micro-cracks on module electric power and the evolution of micro-cracks in long-term cyclic loading remain a topic of ongoing research~\cite{aghaei2022review, silverman2021cracked, Cara2022Crack}, so other factors may also cause the difference in power loss here. Another possible reason is that the current data set may be influenced by other confounding variables. The mitigation of the influence of confounding variables will be illustrated in the following subsection ``Statistical Validation of SHAP Interpretation''. We note that the solar industry is phasing down the usage of poly-c Si cells primarily because of the lowered cost and higher efficiency of mono-c Si cells. The findings from this analysis can provide another motivation for the switch from poly-c to mono-c Si cells. 

\subsubsection{Impact of encapsulant thickness}
The encapsulant thickness remains another important design factor. From Figure~\ref{fig:dependence}(a)(b), both front and rear encapsulant thickness exhibit negative relationship with the Shapley value when the thickness is lower than around $0.6-0.7\text{mm}$, suggesting that thinner encapsulant might result in higher power loss. Particularly, when the encapsulant thickness is reduced to around $0.4\text{mm}$, the influence becomes pronounced, with the front encapsulant increasing the power loss by approximately 0.2\% and the rear encapsulant by approximately 0.6\%. This corresponds to previously simulated results~\cite{bosco2016influence} that thicker encapsulant decreases the accumulated thermal stress in the solder layer. Also, as a soft embedding of silicon solar cells, the encapsulant can compensate for the strain coming from the glass during deflection~\cite{dietrich2012interdependency}. Interestingly, when the front encapsulant thickness is greater than $0.7\text{mm}$, the power loss increases, especially for the front encapsulant. This opposite trend was not reported by previous regression analysis~\cite{park2022reliability, bosco2016influence}. A possible cause for this trend is a change in the mechanical properties of the polymer at lower temperature. At high temperature, the encapsulant layer is soft and acts as a compensation layer for the strain difference between the glass and Si layer. However, lower temperature, especially approaching or below the glass transition temperature ($ \sim -33^\circ\text{C}$~\cite{agroui2012measurement} for EVA and $\sim -25^\circ\text{C}$~\cite{baiamonte2022durability} for POE), limits the mobility of polymer chains and thus increases the stiffness of the encapsulant. This transition transforms the laminated structure of solar module into so-called glass-encapsulant-Si ``sandwich'' structure and more strain is conducted to the cell layer by the stiff encapsulant as the module bends during thermal cycling~\cite{dietrich2012interdependency, dietrich2010mechanical}. In this structure at low temperature, thicker encapsulant between the glass and Si layer, which is the front encapsulant, can increase the tensile stress in the Si layer and lead to higher Si fracture probability. Such a phenomenon was previously reported by Dietrich \emph{et al.}~\cite{dietrich2012interdependency} that, at $-40^\circ\text{C}$, thicker encapsulant leads to higher probability of failure. Therefore, in our BOM data set, the power loss first decreased with thicker encapsulant and then increased as the failure at lower temperature dominates. The hypothesis here can also explain that front encapsulant presents steeper increase at tail region ($>0.7\text{mm}$) in Figure~\ref{fig:dependence}(a) than the rear encapsulant in Figure~\ref{fig:dependence}(b) since the front encapsulant locates between the glass and Si layer conduct more strain. However, we note that since the measurements at the tail region are sparse, this trend of increase may also be influenced by noise. Further testing with more controlled data sets may be needed to validate the hypothesis. 

\subsubsection{Impact of busbar number}
Figure~\ref{fig:dependence}(c) demonstrates that the power loss decreases about 0.6\% as the number of busbars increases from 4 to around 7 and then increases slightly or remains unchanged with more busbars soldered on solar cells. This is reasonable because increasing the number of busbars can increase the probability of current connection between cells and external circuits even when the solder layer beneath some busbars got cracked due to thermal stress. However, too many busbars soldered on the Si wafer can yield more residual thermal stress during the fabrication process, which causes solder disconnection and even wafer fracture during operation~\cite{rendler2016mechanical, shin2018thermal}. Therefore, the accumulated residual stress counteracts the benefit of increasing busbar number. We also notice the drop of Shapley value at 12 busbars but since the measurements are sparse, it cannot represent the general trend of impacts. The relationship between busbar number and power loss in this analysis should attract attention because multi-busbar design is becoming prevalent in recent module manufacturing, but its long-term durability needs further study~\cite{DuramatWebinar}. 

\subsubsection{Impact of wafer thickness}
The wafer thickness was considered as a top important factor in previous studies~\cite{park2022reliability, bosco2016influence} and also remains an important feature in our data set. However, the impact of wafer thickness in our real-world data set shows a different trend to previously simulated results. Figure~\ref{fig:dependence}(d) indicates that a thinner wafer thickness (from $180\text{{\textmu}m}$ to $160\text{{\textmu}m}$) can increase power loss by over 0.7\%, contrary to previous studies that recommended thinner wafers. However, when the thickness is greater than $180\text{{\textmu}m}$, thinner wafer is beneficial for durability. This discrepancy may arise because earlier simulations primarily focused on damage within the solder layer. However, as a brittle material, silicon wafers are more likely to experience catastrophic fractures if the thickness is excessively reduced. This potentially offsets the advantages of reduced damage in the solder layer. This trend aligns with another study showing that higher wafer thickness is important in module design~\cite{beinert2022thermomechanical}.

\subsubsection{Impact of other features}
\begin{figure}
\centering
\includegraphics[width=0.8\textwidth]{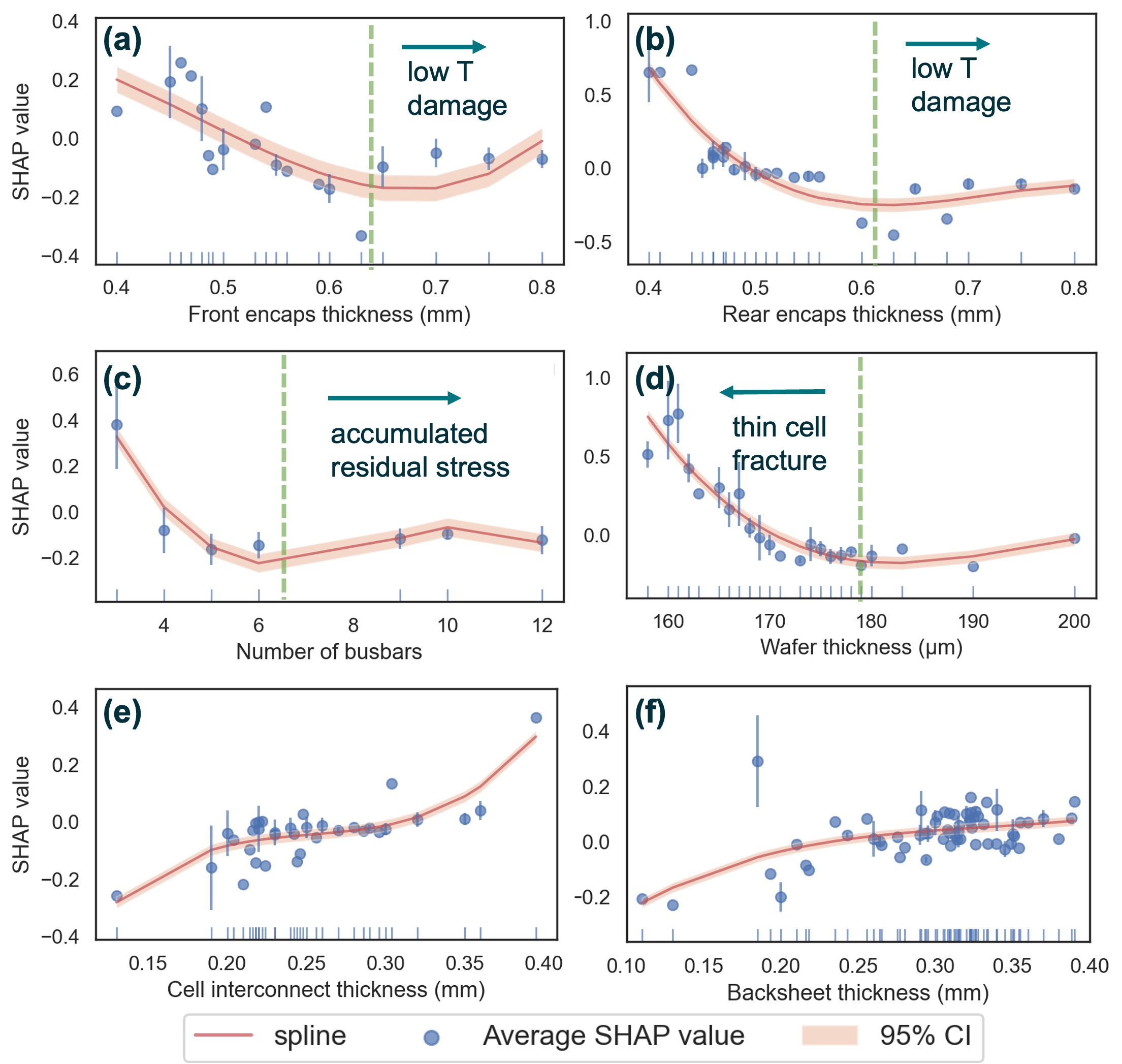}
\caption{Dependence plot of the relation between numerical features and Shapley value. In these dependence plots, the mean Shapley value corresponding to each feature value is displayed as one dot and the standard deviation is used for the error bar. The bottom rug plot in each sub-figure shows the distribution of the data. A smooth spline interpretation with 95\% prediction interval is used to illustrate the trend of the relationship. Due to the additive property of the SHAP formula, the Shapley value is related to the power loss so the dependence plot also reflects the relation between the power loss and the BOM features. (a)-(d) shows the impacts of the top five features apart from the Si cell type. (e) and (f) examine the impact of the two features that were considered in the previous literature.}
\label{fig:dependence}
\end{figure}

Furthermore, Figure~\ref{fig:dependence}(e) and (f) demonstrate the impacts of the cell interconnection (\emph{i.e.,} copper ribbon) thickness and backsheet thickness, which were explored in previous simulations~\cite{park2022reliability, bosco2016influence}. From Figure~\ref{fig:dependence}(e), it can be seen that modules with lower ribbon thickness have lower power loss because of lower thermal stress in the solder layer~\cite{park2022reliability}. However, we consider this trend to be unreliable because the data points in the region over $0.35\text{mm}$ and lower than $0.20\text{mm}$ are sparse and may skew this trend.
Figure~\ref{fig:dependence}(f) suggests that the power loss is expected to increase as the backsheet thickness increases. This increase corresponds to the conclusion of the previous investigation~\cite{bosco2016influence} that thicker backsheet leads to more accumulated damage in the solder layer.  

\subsection{Statistical Validation of SHAP Interpretation}
To validate the results of the aforementioned interpretation, we directly examine the power loss in the original data set. As explained in Section~\ref{methods}, the original data set is not well controlled, so we re-sample a controlled subset derived from the original one. To determine whether Si type is indeed impactful, we construct this subset using k-means clustering method so that feature values in this subset are similar apart from the Si type. Then we conduct a t-test on this subset and illustrate that poly-c Si modules have a statistically significant increase in the power loss. In this section we select the Si type for this analysis because not only is it the most impactful factor, but it is a categorical variable which makes the control process easier. The example of clustering other numerical features is included in the SI Section ``Post hoc Statistical Analysis'', which shows that the control group is hard to obtain for other variables.

\begin{figure*}
\centering
\includegraphics[width=0.9\textwidth]{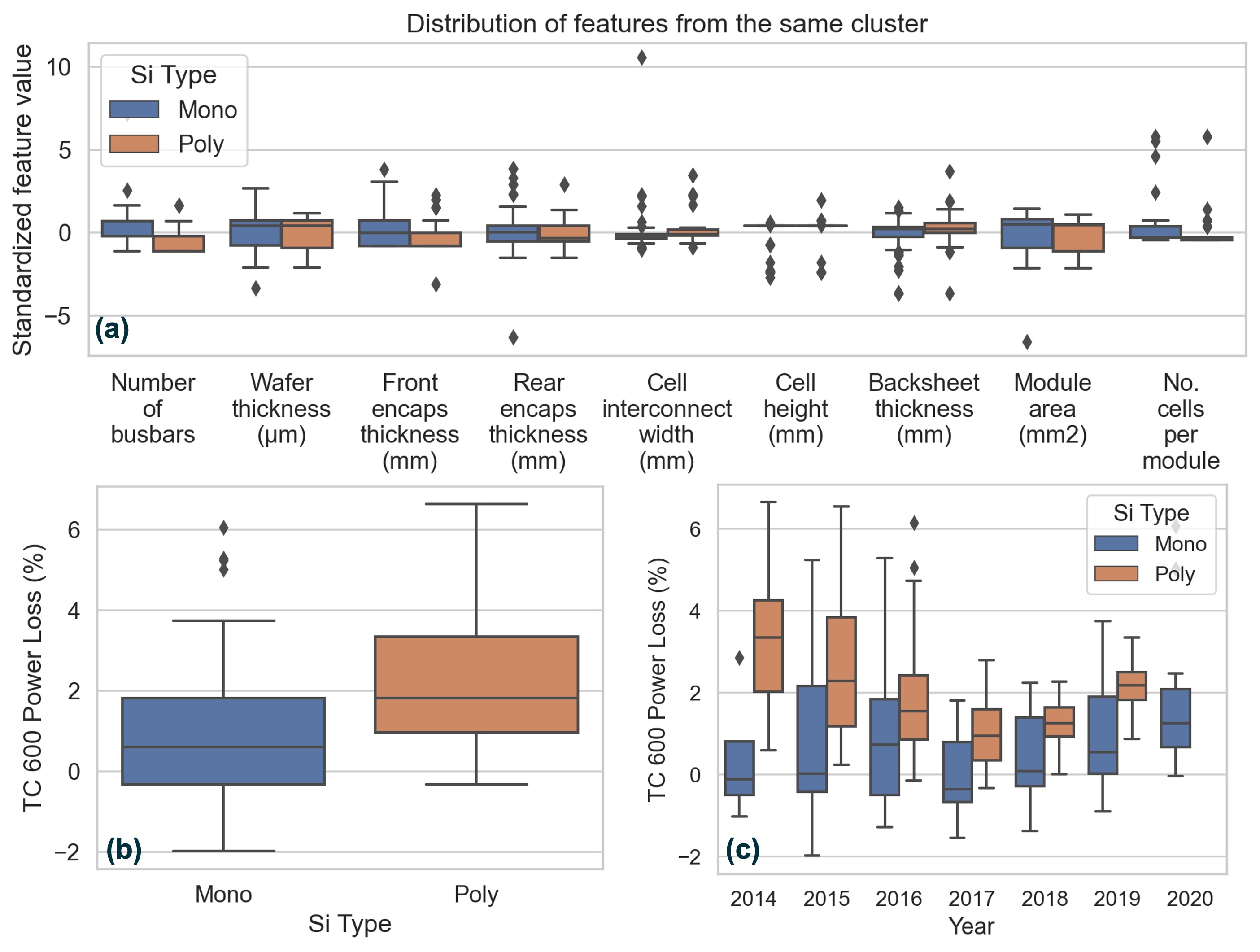}
\caption{Statistical testing to verify the impact of the Si cell type. Sample data is first extracted from the raw data set using the clustering method to guarantee that features other than Si cell type are randomized. Then the t-test is used to test the statistical significance of the impact of the Si cell type. (a) The distribution of several top important features in the same cluster. The similar distribution comparison between mono-c Si cells and poly-c Si cells indicates the randomization of these features. (b) The distribution of power loss between mono-c Si cells and poly-c Si cells in the same cluster. The p-value smaller than 5\% indicates that poly-c Si modules have a statistically significant increase in the thermal cycling power loss. (c) The distribution of Si type in each year. The higher power loss of poly-c Si modules indicates that the manufacturing year is not the confounding variable.}
\label{fig:t-test}
\end{figure*}

In this controlled subset to test the impact of the Si type, the distribution of the top ten important BOM features other than Si type is shown in Figure~\ref{fig:t-test}(a). It can be seen that the boxplots of most features for these cells are similar but contain differences in Si type, which means that these features are randomized between mono-c and poly-c Si. We notice that poly-c Si cells tend to have fewer busbars, which may cause a slight correlation between ``Si type'' and ``number of busbars''. This potential correlation may influence the importance of these two features. However, after checking the original data set, we find that the difference of median value of the number of busbars is not large between mono-c (median=5) and poly-c Si cells (median=4), so the influence of this difference may not be significant since the difference of Shapley value in 4 and 5 busbars is around 0.1. More details of constructing this cluster is shown in the SI section ``Post hoc Statistical Analysis''. We also check the normality of the power loss using the quantile-quantile (Q-Q) plots, which is a prerequisite for further parametric statistical testing. The details are included in the SI. 

With all prerequisite processing of the data set, we conduct a t-test to compare the mean power loss of poly-c Si modules and mono-c Si modules in this controlled subset. Figure~\ref{fig:t-test}(b) compares the distribution of the power loss of mono-c and poly-c Si modules from the selected cluster, which displays that poly-c Si modules have more power loss on average. The null hypothesis of our t-test is:
\begin{flalign*}
\quad\quad&\mathbf{H_0}: \textit{Poly-c Si modules don't exhibit higher thermal cycling power loss than mono-c Si modules.} 
\end{flalign*}

The test result illustrates that the power loss of poly-c Si modules is greater than that of mono-c Si modules and this result is statistically significant ($\text{p-value}=1.4\times10^{-8}<5\%$). This validates the interpretation of the SHAP analysis.

Although the clustering can mitigate the influence of uncontrolled variables, some other variables that are not included in the ML modeling may still impact our interpretation. One of the most potential confounding variables is the manufacturing process. To mitigate the bias towards a specific supplier, this data set collects modules from various manufacturers. Also in Figure~\ref{fig:t-test}(c), we compare the power loss of the two types of modules from the same cluster in each manufacturing year. It reveals that poly-c Si modules still have a higher power loss each year, which means that the manufacturing year is not the underlying reason for the difference. Despite these efforts, there is still a chance that some unrecorded manufacturing conditions during the fabrication of poly-c and mono-c Si modules may vary and influence the thermomechanical durability rather than the Si type itself. As in all such cases, data analysis can point the direction to highly plausible hypotheses, but careful and dedicated experiments would be needed to confirm each trend.

\section{Conclusion} \label{conclusion}
In this work, we identify that the predominant BOM features related to module's thermomechanical durability in the current module design are Si type, encapsulant thickness, busbar numbers, and wafer thickness. Our analysis is based on a unique data set that includes the BOM features of full-size modules from various manufacturers and their power loss after 600 thermal cycles. To analyze this data set, we first correlate the BOM features with the power loss using machine learning modeling. By comparing with other models, we develop a RF model with a testing RMSE of 1.179\% as the optimal model and a RMSE of 0.558\% on the whole data set for further SHAP interpretation. We subsequently apply SHAP analysis to the fitted RF model to interpret the whole data set and used statistical testing to verify the conclusion. 

Overall, we find that the Si type (\emph{i.e.,} whether the module is composed of poly-c or mono-c Si cells) is the most influential factor. The replacement of poly-c Si with mono-c Si decreases the power loss by approximately 0.6\% in the SHAP analysis. The next most important feature is the thickness of the front and rear encapsulant; a higher thickness can reduce power loss, but further increases in thickness (over around $0.6-0.7\text{mm}$) present an opposite impact, particularly for the front encapsulant. We also analyze the impacts of other features. In particular, we find that increasing the number of busbars initially reduces power loss but subsequently increases it after the number exceeds around 7. The wafer thickness displays an opposite trend to the previous findings~\cite{park2022reliability, bosco2016influence}. In our data set, too low wafer thickness ($<180\text{{\textmu}m}$) is not beneficial for the durability. We further find that the power loss is reduced when the ribbon thickness is decreased and thinner backsheet layer is beneficial for durability. 

Finally, it is important to acknowledge the limitations of our study. Although our data set covers a breadth of BOM features, certain areas, such as the solder thickness or various CTEs of encapsulants, remained uncharted due to the difficulty of tracking this information from manufacturers. Furthermore, the use of commercially available modules as opposed to carefully controlled modules means that conclusions might still be influenced by confounding variables, underscoring the need for further investigation. Despite these constraints, we hope this study can reveal the potential application of machine learning and model-agnostic interpretation methods in examining BOM effects on reliability and provide insight into the direction of future module optimization. 

\section{Data Availability}
The data set is available at DuraMat Datahub (\url{https://datahub.duramat.org/dataset/bom_thermal_cycling_degradation}). Some values in the published data set are redacted according to non-disclosure agreement.

\section{Code Availability}
Our codes for ML modeling and SHAP interpretation are available at Github repository(\url{https://github.com/DuraMAT/bom_analysis}).

\section*{Acknowledgment}
The authors would like to thank PVEL for providing the data. Especially the authors would like to thank Tristan Erion-Lorico and Max Macpherson from PVEL for coordinating this work. The authors also appreciate feedback on this work from Nick Bosco, Baojie Li, and Zhuoying Zhu.

\section*{Declaration of Generative AI and AI-assisted technologies in the writing process}
During the preparation of this work the author(s) used Writefull \& Chatgpt in order to check grammar and improve readability. The author(s) did not use this tool/service to directly generate the manuscript but only to polish the manuscript written by the author(s). After using this tool/service, the author(s) reviewed and edited the content as needed and take(s) full responsibility for the content of the publication.
\bibliographystyle{IEEEtran}
\bibliography{refs.bib}

\begin{thebibliography}{10}
\providecommand{\url}[1]{#1}
\csname url@samestyle\endcsname
\providecommand{\newblock}{\relax}
\providecommand{\bibinfo}[2]{#2}
\providecommand{\BIBentrySTDinterwordspacing}{\spaceskip=0pt\relax}
\providecommand{\BIBentryALTinterwordstretchfactor}{4}
\providecommand{\BIBentryALTinterwordspacing}{\spaceskip=\fontdimen2\font plus
\BIBentryALTinterwordstretchfactor\fontdimen3\font minus \fontdimen4\font\relax}
\providecommand{\BIBforeignlanguage}[2]{{%
\expandafter\ifx\csname l@#1\endcsname\relax
\typeout{** WARNING: IEEEtran.bst: No hyphenation pattern has been}%
\typeout{** loaded for the language `#1'. Using the pattern for}%
\typeout{** the default language instead.}%
\else
\language=\csname l@#1\endcsname
\fi
#2}}
\providecommand{\BIBdecl}{\relax}
\BIBdecl

\bibitem{Ardani2022solar}
K.~Ardani, P.~Denholm, T.~Mai, R.~Margolis, E.~O'Shaughnessy, T.~Silverman, and J.~Zuboy, ``Solar futures study,'' U.S. Department of Energy, Tech. Rep., 2021.

\bibitem{aghaei2022review}
M.~Aghaei, A.~Fairbrother, A.~Gok, S.~Ahmad, S.~Kazim, K.~Lobato, G.~Oreski, A.~Reinders, J.~Schmitz, M.~Theelen \emph{et~al.}, ``Review of degradation and failure phenomena in photovoltaic modules,'' \emph{Renewable and Sustainable Energy Reviews}, vol. 159, p. 112160, 2022.

\bibitem{jordan2017photovoltaic}
D.~C. Jordan, T.~J. Silverman, J.~H. Wohlgemuth, S.~R. Kurtz, and K.~T. VanSant, ``Photovoltaic failure and degradation modes,'' \emph{Progress in Photovoltaics: Research and Applications}, vol.~25, no.~4, pp. 318--326, 2017.

\bibitem{kawai2017causes}
S.~Kawai, T.~Tanahashi, Y.~Fukumoto, F.~Tamai, A.~Masuda, and M.~Kondo, ``Causes of degradation identified by the extended thermal cycling test on commercially available crystalline silicon photovoltaic modules,'' \emph{IEEE Journal of Photovoltaics}, vol.~7, no.~6, pp. 1511--1518, 2017.

\bibitem{eitner2010use}
U.~Eitner, M.~K{\"o}ntges, and R.~Brendel, ``Use of digital image correlation technique to determine thermomechanical deformations in photovoltaic laminates: Measurements and accuracy,'' \emph{Solar Energy Materials and Solar Cells}, vol.~94, no.~8, pp. 1346--1351, 2010.

\bibitem{borri2018fatigue}
C.~Borri, M.~Gagliardi, and M.~Paggi, ``Fatigue crack growth in silicon solar cells and hysteretic behaviour of busbars,'' \emph{Solar Energy Materials and Solar Cells}, vol. 181, pp. 21--29, 2018.

\bibitem{wiese2010mechanical}
S.~Wiese, R.~Meier, and F.~Kraemer, ``Mechanical behaviour and fatigue of copper ribbons used as solar cell interconnectors,'' in \emph{2010 11th International Thermal, Mechanical \& Multi-Physics Simulation, and Experiments in Microelectronics and Microsystems (EuroSimE)}.\hskip 1em plus 0.5em minus 0.4em\relax IEEE, 2010, pp. 1--5.

\bibitem{silverman2019movement}
T.~J. Silverman, M.~Bliss, A.~Abbas, T.~Betts, M.~Walls, and I.~Repins, ``Movement of cracked silicon solar cells during module temperature changes,'' in \emph{2019 IEEE 46th Photovoltaic Specialists Conference (PVSC)}.\hskip 1em plus 0.5em minus 0.4em\relax IEEE, 2019, pp. 1517--1520.

\bibitem{park2022reliability}
S.~Park and C.~Han, ``Reliability-driven design optimization of si solar module under thermal cycling,'' \emph{Journal of Mechanical Science and Technology}, vol.~36, no.~8, pp. 4099--4114, 2022.

\bibitem{jeong2012field}
J.-S. Jeong, N.~Park, and C.~Han, ``Field failure mechanism study of solder interconnection for crystalline silicon photovoltaic module,'' \emph{Microelectronics Reliability}, vol.~52, no. 9-10, pp. 2326--2330, 2012.

\bibitem{itoh2014solder}
U.~Itoh, M.~Yoshida, H.~Tokuhisa, K.~Takeuchi, and Y.~Takemura, ``Solder joint failure modes in the conventional crystalline si module,'' \emph{Energy Procedia}, vol.~55, pp. 464--468, 2014.

\bibitem{hanifi2020loss}
H.~Hanifi, M.~Pander, U.~Zeller, K.~Ilse, D.~Dassler, M.~Mirza, M.~A. Bahattab, B.~Jaeckel, C.~Hagendorf, M.~Ebert \emph{et~al.}, ``Loss analysis and optimization of pv module components and design to achieve higher energy yield and longer service life in desert regions,'' \emph{Applied energy}, vol. 280, p. 116028, 2020.

\bibitem{bosco2016influence}
N.~Bosco, T.~J. Silverman, and S.~Kurtz, ``The influence of pv module materials and design on solder joint thermal fatigue durability,'' \emph{IEEE Journal of Photovoltaics}, vol.~6, no.~6, pp. 1407--1412, 2016.

\bibitem{zhu2018effect}
J.~Zhu, M.~Owen-Bellini, D.~Montiel-Chicharro, T.~R. Betts, and R.~Gottschalg, ``Effect of viscoelasticity of ethylene vinyl acetate encapsulants on photovoltaic module solder joint degradation due to thermomechanical fatigue,'' \emph{Japanese Journal of Applied Physics}, vol.~57, no. 8S3, p. 08RG03, 2018.

\bibitem{beinert2019effect}
A.~J. Beinert, P.~Romer, M.~Heinrich, M.~Mittag, J.~Aktaa, and D.~H. Neuhaus, ``The effect of cell and module dimensions on thermomechanical stress in pv modules,'' \emph{IEEE Journal of Photovoltaics}, vol.~10, no.~1, pp. 70--77, 2019.

\bibitem{beinert2022thermomechanical}
A.~J. Beinert, P.~Romer, M.~Heinrich, J.~Aktaa, and H.~Neuhaus, ``Thermomechanical design rules for photovoltaic modules,'' \emph{Progress in Photovoltaics: Research and Applications}, 2022.

\bibitem{friedman2001greedy}
J.~H. Friedman, ``Greedy function approximation: a gradient boosting machine,'' \emph{Annals of statistics}, pp. 1189--1232, 2001.

\bibitem{ribeiro2016should}
M.~T. Ribeiro, S.~Singh, and C.~Guestrin, ``" why should i trust you?" explaining the predictions of any classifier,'' in \emph{Proceedings of the 22nd ACM SIGKDD international conference on knowledge discovery and data mining}, 2016, pp. 1135--1144.

\bibitem{lundberg2017unified}
S.~M. Lundberg and S.-I. Lee, ``A unified approach to interpreting model predictions,'' \emph{Advances in neural information processing systems}, vol.~30, 2017.

\bibitem{bannigan2023machine}
P.~Bannigan, Z.~Bao, R.~J. Hickman, M.~Aldeghi, F.~H{\"a}se, A.~Aspuru-Guzik, and C.~Allen, ``Machine learning models to accelerate the design of polymeric long-acting injectables,'' \emph{Nature communications}, vol.~14, no.~1, p.~35, 2023.

\bibitem{giles2022machine}
S.~A. Giles, D.~Sengupta, S.~R. Broderick, and K.~Rajan, ``Machine-learning-based intelligent framework for discovering refractory high-entropy alloys with improved high-temperature yield strength,'' \emph{npj Computational Materials}, vol.~8, no.~1, p. 235, 2022.

\bibitem{kumar2021chemical}
R.~Kumar and A.~K. Singh, ``Chemical hardness-driven interpretable machine learning approach for rapid search of photocatalysts,'' \emph{npj Computational Materials}, vol.~7, no.~1, p. 197, 2021.

\bibitem{hartono2020machine}
N.~T.~P. Hartono, J.~Thapa, A.~Tiihonen, F.~Oviedo, C.~Batali, J.~J. Yoo, Z.~Liu, R.~Li, D.~F. Marr{\'o}n, M.~G. Bawendi \emph{et~al.}, ``How machine learning can help select capping layers to suppress perovskite degradation,'' \emph{Nature communications}, vol.~11, no.~1, p. 4172, 2020.

\bibitem{international2016photovoltaic}
I.~E. Commission, ``Terrestrial photovoltaic (pv) modules - design qualification and type approval - part 2: Test procedures,'' International Electrotechnical Commission, Tech. Rep., 2016.

\bibitem{ho1995random}
T.~K. Ho, ``Random decision forests,'' in \emph{Proceedings of 3rd international conference on document analysis and recognition}, vol.~1.\hskip 1em plus 0.5em minus 0.4em\relax IEEE, 1995, pp. 278--282.

\bibitem{pvel}
\BIBentryALTinterwordspacing
Pv evolution labs. [Online]. Available: \url{https://www.pvel.com/}
\BIBentrySTDinterwordspacing

\bibitem{python3}
G.~Van~Rossum and F.~L. Drake, \emph{Python 3 Reference Manual}.\hskip 1em plus 0.5em minus 0.4em\relax Scotts Valley, CA: CreateSpace, 2009.

\bibitem{PVELcard}
\BIBentryALTinterwordspacing
PVEL, ``The 2023 pv module reliability scorecard,'' 2023. [Online]. Available: \url{https://scorecard.pvel.com/}
\BIBentrySTDinterwordspacing

\bibitem{international2006photovoltaic}
I.~E. Commission, ``Photovoltaic devices – part 1: Measurement of photovoltaic current-voltage characteristics,'' International Electrotechnical Commission, Tech. Rep., 2006.

\bibitem{troyanskaya2001missing}
O.~Troyanskaya, M.~Cantor, G.~Sherlock, P.~Brown, T.~Hastie, R.~Tibshirani, D.~Botstein, and R.~B. Altman, ``Missing value estimation methods for dna microarrays,'' \emph{Bioinformatics}, vol.~17, no.~6, pp. 520--525, 2001.

\bibitem{vallat2018pingouin}
R.~Vallat, ``Pingouin: statistics in python.'' \emph{J. Open Source Softw.}, vol.~3, no.~31, p. 1026, 2018.

\bibitem{2020SciPy-NMeth}
P.~Virtanen, R.~Gommers, T.~E. Oliphant, M.~Haberland, T.~Reddy, D.~Cournapeau, E.~Burovski, P.~Peterson, W.~Weckesser, J.~Bright, S.~J. {van der Walt}, M.~Brett, J.~Wilson, K.~J. Millman, N.~Mayorov, A.~R.~J. Nelson, E.~Jones, R.~Kern, E.~Larson, C.~J. Carey, {\.I}.~Polat, Y.~Feng, E.~W. Moore, J.~{VanderPlas}, D.~Laxalde, J.~Perktold, R.~Cimrman, I.~Henriksen, E.~A. Quintero, C.~R. Harris, A.~M. Archibald, A.~H. Ribeiro, F.~Pedregosa, P.~{van Mulbregt}, and {SciPy 1.0 Contributors}, ``{{SciPy} 1.0: Fundamental Algorithms for Scientific Computing in Python},'' \emph{Nature Methods}, vol.~17, pp. 261--272, 2020.

\bibitem{bruce2020practical}
P.~Bruce, A.~Bruce, and P.~Gedeck, \emph{Practical statistics for data scientists: 50+ essential concepts using R and Python}.\hskip 1em plus 0.5em minus 0.4em\relax O'Reilly Media, 2020.

\bibitem{scikit-learn}
F.~Pedregosa, G.~Varoquaux, A.~Gramfort, V.~Michel, B.~Thirion, O.~Grisel, M.~Blondel, P.~Prettenhofer, R.~Weiss, V.~Dubourg, J.~Vanderplas, A.~Passos, D.~Cournapeau, M.~Brucher, M.~Perrot, and E.~Duchesnay, ``Scikit-learn: Machine learning in {P}ython,'' \emph{Journal of Machine Learning Research}, vol.~12, pp. 2825--2830, 2011.

\bibitem{chen2016xgboost}
T.~Chen and C.~Guestrin, ``Xgboost: A scalable tree boosting system,'' in \emph{Proceedings of the 22nd acm sigkdd international conference on knowledge discovery and data mining}, 2016, pp. 785--794.

\bibitem{tibshirani1996regression}
R.~Tibshirani, ``Regression shrinkage and selection via the lasso,'' \emph{Journal of the Royal Statistical Society Series B: Statistical Methodology}, vol.~58, no.~1, pp. 267--288, 1996.

\bibitem{hoerl1970ridge}
A.~E. Hoerl and R.~W. Kennard, ``Ridge regression: Biased estimation for nonorthogonal problems,'' \emph{Technometrics}, vol.~12, no.~1, pp. 55--67, 1970.

\bibitem{drucker1996support}
H.~Drucker, C.~J. Burges, L.~Kaufman, A.~Smola, and V.~Vapnik, ``Support vector regression machines,'' \emph{Advances in neural information processing systems}, vol.~9, 1996.

\bibitem{efron1992bootstrap}
B.~Efron, ``Bootstrap methods: another look at the jackknife,'' in \emph{Breakthroughs in statistics: Methodology and distribution}.\hskip 1em plus 0.5em minus 0.4em\relax Springer, 1992, pp. 569--593.

\bibitem{kmeans}
J.~A. Hartigan and M.~A. Wong, ``Algorithm as 136: A k-means clustering algorithm,'' \emph{Journal of the royal statistical society. series c (applied statistics)}, vol.~28, no.~1, pp. 100--108, 1979.

\bibitem{watanabe2004linear}
H.~Watanabe, N.~Yamada, and M.~Okaji, ``Linear thermal expansion coefficient of silicon from 293 to 1000 k,'' \emph{International journal of thermophysics}, vol.~25, no.~1, pp. 221--236, 2004.

\bibitem{zavattieri2001grain}
P.~D. Zavattieri and H.~D. Espinosa, ``Grain level analysis of crack initiation and propagation in brittle materials,'' \emph{Acta Materialia}, vol.~49, no.~20, pp. 4291--4311, 2001.

\bibitem{wu2013effect}
H.~Wu and S.~N. Melkote, ``Effect of crystal defects on mechanical properties relevant to cutting of multicrystalline solar silicon,'' \emph{Materials science in semiconductor processing}, vol.~16, no.~6, pp. 1416--1421, 2013.

\bibitem{silverman2021cracked}
T.~J. Silverman, M.~G. Deceglie, M.~Owen-Bellini, W.~B. Hobbs, and C.~Libby, ``Cracked solar cell performance depends on module temperature,'' in \emph{2021 IEEE 48th Photovoltaic Specialists Conference (PVSC)}.\hskip 1em plus 0.5em minus 0.4em\relax IEEE, 2021, pp. 1691--1692.

\bibitem{Cara2022Crack}
C.~Libby, B.~Paudyal, X.~Chen, W.~B. Hobbs, D.~Fregosi, and A.~Jain, ``Analysis of pv module power loss and cell crack effects due to accelerated aging tests and field exposure,'' \emph{IEEE Journal of Photovoltaics}, pp. 1--9, 2022.

\bibitem{dietrich2012interdependency}
S.~Dietrich, M.~Sander, M.~Pander, and M.~Ebert, ``Interdependency of mechanical failure rate of encapsulated solar cells and module design parameters,'' in \emph{Reliability of Photovoltaic Cells, Modules, Components, and Systems V}, vol. 8472.\hskip 1em plus 0.5em minus 0.4em\relax SPIE, 2012, pp. 123--131.

\bibitem{agroui2012measurement}
K.~Agroui, G.~Collins, and J.~Farenc, ``Measurement of glass transition temperature of crosslinked eva encapsulant by thermal analysis for photovoltaic application,'' \emph{Renewable Energy}, vol.~43, pp. 218--223, 2012.

\bibitem{baiamonte2022durability}
M.~Baiamonte, C.~Colletti, A.~Ragonesi, C.~Gerardi, and N.~T. Dintcheva, ``Durability and performance of encapsulant films for bifacial heterojunction photovoltaic modules,'' \emph{Polymers}, vol.~14, no.~5, p. 1052, 2022.

\bibitem{dietrich2010mechanical}
S.~Dietrich, M.~Pander, M.~Sander, S.~H. Schulze, and M.~Ebert, ``Mechanical and thermomechanical assessment of encapsulated solar cells by finite-element-simulation,'' in \emph{Reliability of photovoltaic cells, modules, components, and systems III}, vol. 7773.\hskip 1em plus 0.5em minus 0.4em\relax SPIE, 2010, pp. 117--126.

\bibitem{rendler2016mechanical}
L.~C. Rendler, A.~Kraft, C.~Ebert, U.~Eitner, and S.~Wiese, ``Mechanical stress in solar cells with multi busbar interconnection—parameter study by fem simulation,'' in \emph{2016 17th International Conference on Thermal, Mechanical and Multi-Physics Simulation and Experiments in Microelectronics and Microsystems (EuroSimE)}.\hskip 1em plus 0.5em minus 0.4em\relax IEEE, 2016, pp. 1--5.

\bibitem{shin2018thermal}
H.~Shin, E.~Han, N.~Park, and D.~Kim, ``Thermal residual stress analysis of soldering and lamination processes for fabrication of crystalline silicon photovoltaic modules,'' \emph{Energies}, vol.~11, no.~12, p. 3256, 2018.

\bibitem{DuramatWebinar}
\BIBentryALTinterwordspacing
J.~Zuboy, M.~Springer, E.~Palmiotti, T.~Barnes, J.~Karas, B.~Smith, and M.~Woodhouse. (2022) Technology scouting report: Reliability implications of recent pv module technology trends. PDF document. [Online]. Available: \url{https://www.nrel.gov/docs/fy22osti/82871.pdf}
\BIBentrySTDinterwordspacing

\end{thebibliography}
\end{document}


\title{Supplementary Information: Analyzing the Impact of Design Factors on Solar Module Thermomechanical Durability Using Interpretable Machine Learning Techniques}

\author{
\IEEEauthorblockN{Xin Chen~\IEEEauthorrefmark{1}\IEEEauthorrefmark{2}, Todd Karin~\IEEEauthorrefmark{3}, Anubhav Jain~\IEEEauthorrefmark{1} \vspace{15pt}} \\
\IEEEauthorblockA{
\IEEEauthorrefmark{1}Lawrence Berkeley National Laboratory, Berkeley, CA, U.S.A \\
\IEEEauthorrefmark{2}University of California, Berkeley, Berkeley, CA, U.S.A \\
\IEEEauthorrefmark{3}Kiwa PVEL, Member of Kiwa Group, Napa, CA, U.S.A
}}

\maketitle

\section{Data Preprocessing}
\subsection{Detecting Outliers}
\begin{figure}[H]
\centering
\includegraphics[width=0.6\textwidth]{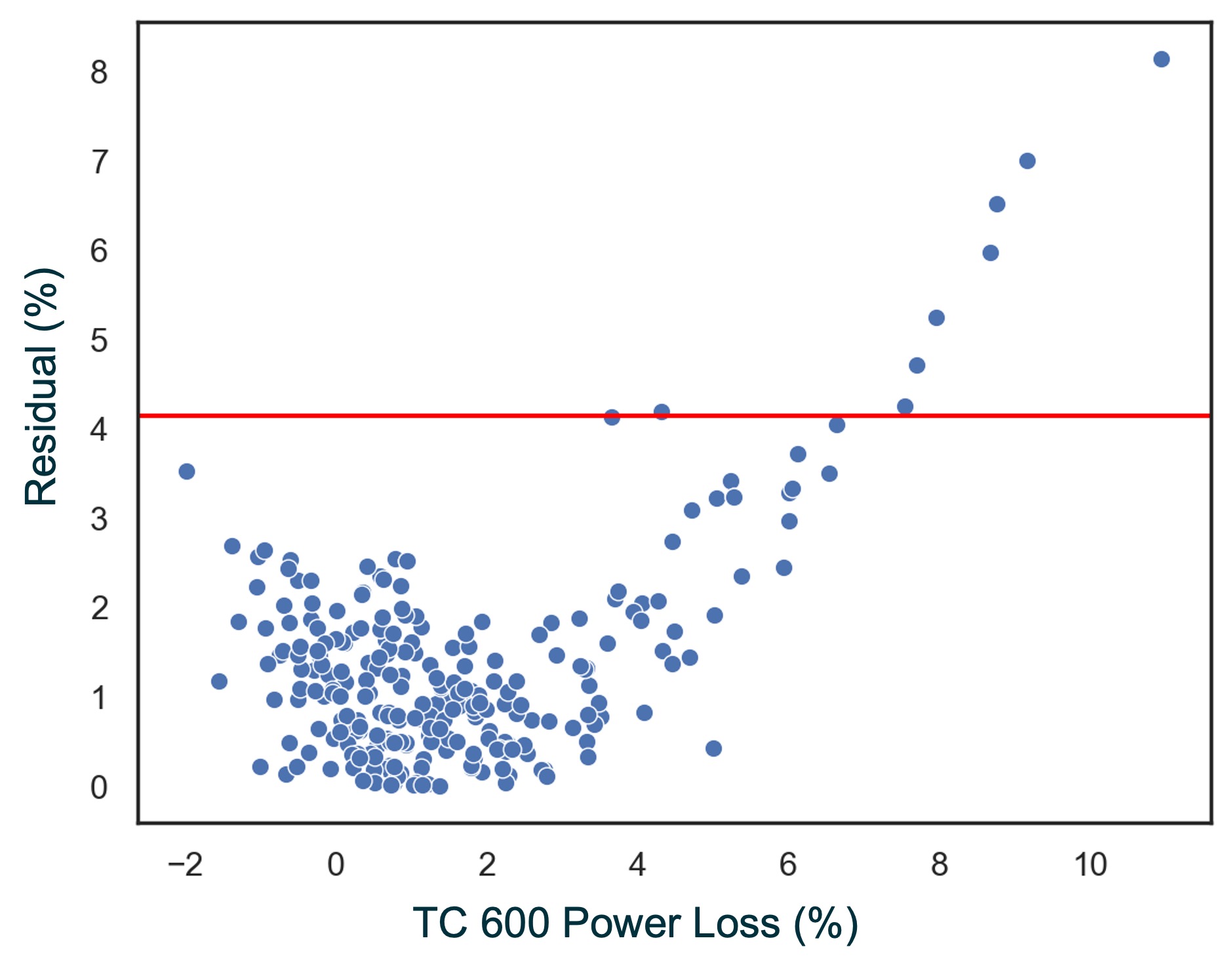}
\caption{Residual analysis to detect outliers. The red line indicates the 97th percentile.}
\label{fig:outlier}
\end{figure}

We filter outliers such as censored data or extreme values by doing both residual analysis and manual inspection. When doing residual analysis, we first fit a linear model (ordinary least square) on the whole data set. The residual is calculated as Equation~\ref{residual}. Then we filter out the observations that yield residuals in the 97th percentile or higher of the residual distribution, as shown in Figure~\ref{fig:outlier}. Those outliers turn out to be extreme values far away from the central distribution. It should be noted that this doesn't necessarily mean the measurement of those modules is invalid, but only indicates that those data points may decrease model performance. Also, the selection of the threshold of filtering is subjective. Since the amount of all the data is not high enough, 97th percentile is selected to remove outliers in the tail region but also keep enough data for modeling.

\begin{equation}
\text{residual}=\left| \text{prediction} - \text{measurement} \right|
\label{residual}
\end{equation}

\subsection{Transforming Records and Data Type Casting}
We fix wrong records and transform ununified data types in this step. For example, some data points record the module width and height in the opposite order and must be reversed. Also, numerical values like wafer thickness should have a ``float'' data type, but some records are ``string'' type like ``180 (mm)''. Thus the data type needs to be unified. Also, categorical data is transformed using one-hot encoding. 
\subsection{Handling Missing Data}
We process missing values by imputation or discarding if a large portion of the values are missed. The K nearest neighbor imputer (KNN)~\cite{troyanskaya2001missing} implemented with Scikit-learn is used to fill in the missing numerical values in the data set. The number of neighbors is set as 5. We note that we use more variables presented in the original BOM dataset than in the feature matrix constructed after feature selection. An important variable we included in the KNN model is the manufacturer name, as different sources of manufacturing process may influence the module design and quality. A simple imputer is utilized to fill in the missing categorical data using the most frequent values. We note that other confounding variables related to manufacturing process may influence the KNN performance

\section{Feature Selection}
\subsection{Formulae for Testing Correlation and Statistical Association}

The Spearman correlation is used to test the correlation between two numerical variables, following Equation~\ref{spearman}.
\begin{equation}
\rho_{\mathrm{R}(X), \mathrm{R}(Y)}=\frac{\operatorname{cov}(\mathrm{R}(X), \mathrm{R}(Y))}{\sigma_{\mathrm{R}(X)} \sigma_{\mathrm{R}(Y)}}
\label{spearman}
\end{equation}
where $\operatorname{cov}(\mathrm{R}(X), \mathrm{R}(Y))$ is the covariance of the ranked variables. $\sigma_{\mathrm{R}(X)}$ and $\sigma_{\mathrm{R}(Y)}$ are the standard deviations of the rank variables. Analysis of variance (ANOVA) is used to test the correlation between a numerical variable and a categorical variable following Equation~\ref{anova}
\begin{equation}
\begin{split}
F & = \frac{\text{between-group variance}}{\text{within-group variance}} \\
& = \frac{\sum_{i=1}^K n_i\left(\bar{Y}_i-\bar{Y}\right)^2 /(K-1)}{\sum_{i=1}^K \sum_{j=1}^{n_i}\left(Y_{i j}-\bar{Y}_i\right)^2 /(N-K)}
\end{split}
\label{anova}
\end{equation}
where $\bar{Y}_i$ denotes the sample mean in the $i$-th group, $Y_{i j}$ is the $j$-th observation in the $i$-th out of groups, $\bar{Y}$ denotes the overall mean of the data, $n_{i}$ is the number of observations in the $i$-th group, $N$ is the overall sample size and $K$ denotes the number of groups. Chi-square was used to test the correlation between two categorical variables, following Equation~\ref{chi}.
\begin{equation}
\chi^2=\sum \frac{\left(O_i-E_i\right)^2}{E_i}
\label{chi}
\end{equation}
where $O_i$ is the observed value and $E_i$ is the expected value. A p-value smaller than 5\% indicates a statistical association in both ANOVA and Chi-square tests.

\subsection{Correlation Matrix and Statistical Association Matrix}
\begin{figure}[H]
\centering
\includegraphics[width=0.9\textwidth]{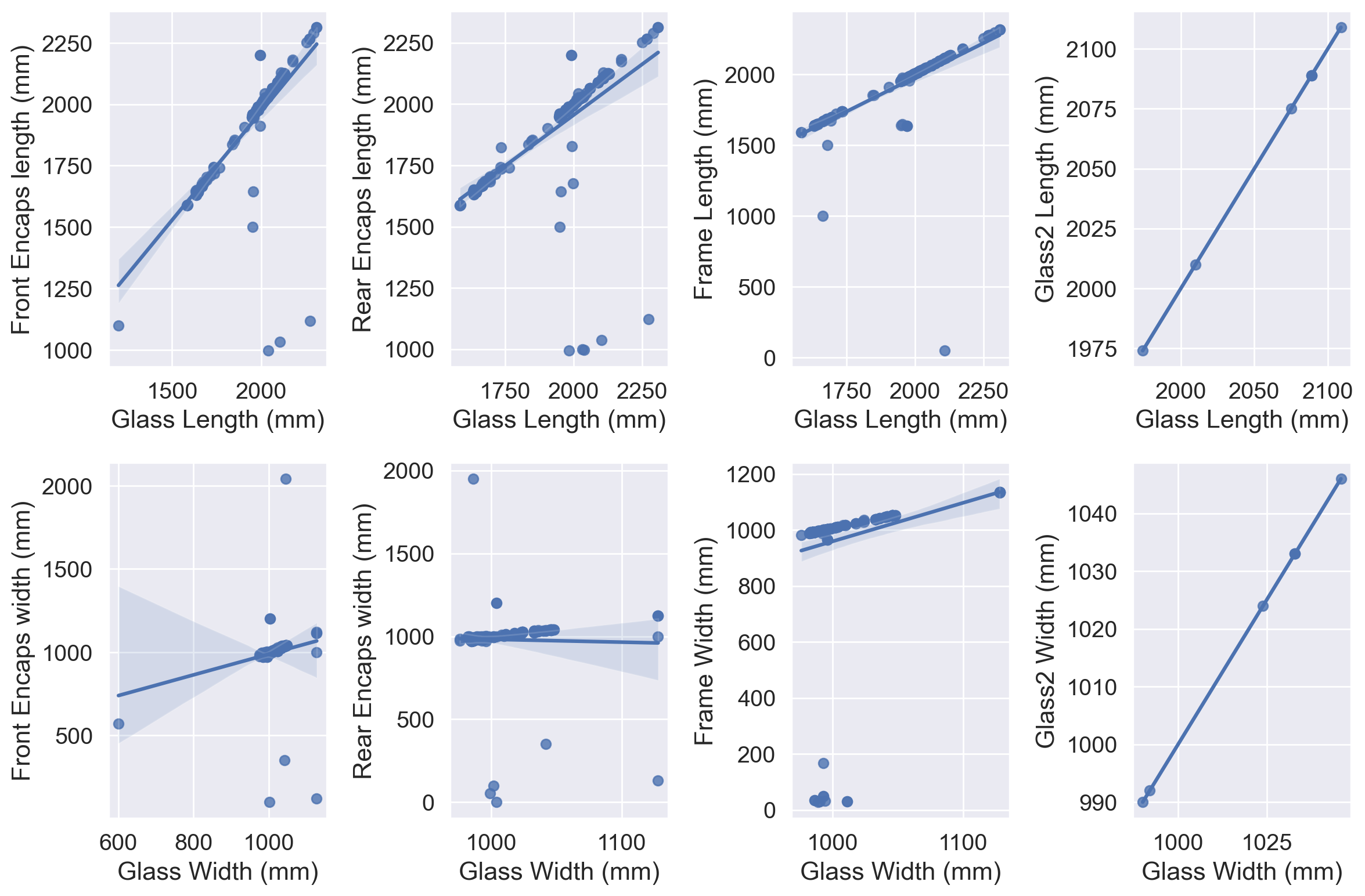}
\caption{Correlation of dimensions between the front glass and other BOM features.}
\label{fig:glass-corr}
\end{figure}

Figure~\ref{fig:glass-corr} illustrates strong correlations between the dimensions (length and width) of the front glass and other module layers. This means in practical designs, modules with larger front glass dimensions typically have similarly sized other layers. Consequently, we consider these dimensions as identical variables and retain only the length and width of the front glass in the feature matrix. This helps to prevent lower feature importance in the presence of dependent features. For instance, the front encapsulant is closely correlated with the length of the glass as shown in Figure~\ref{fig:glass-corr}. The model can predict using either of these, which dilutes the importance of both features. Removing other dimensions from the feature matrix ensures that the glass dimensions alone adequately represent the importance of these layers in SHAP analysis. 

\begin{figure}[H]
\centering
\includegraphics[width=0.9\textwidth]{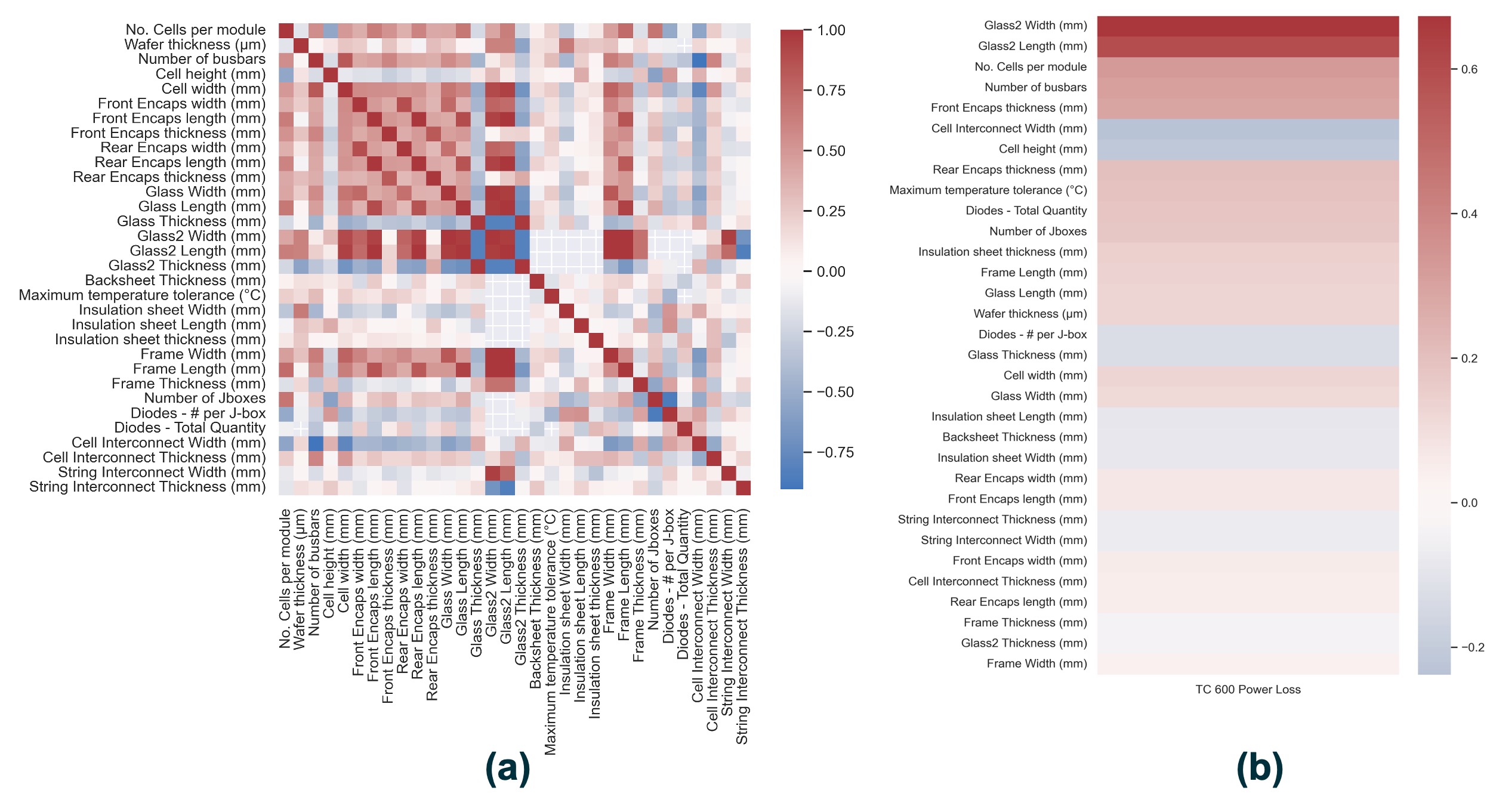}
\caption{Full correlation matrix among numerical features and the target variable. (a) The Spearman correlation among features. (b) The correlation between features and target variable.}
\label{fig:spear-corr}
\end{figure}

\begin{figure}[H]
\centering
\includegraphics[width=0.9\textwidth]{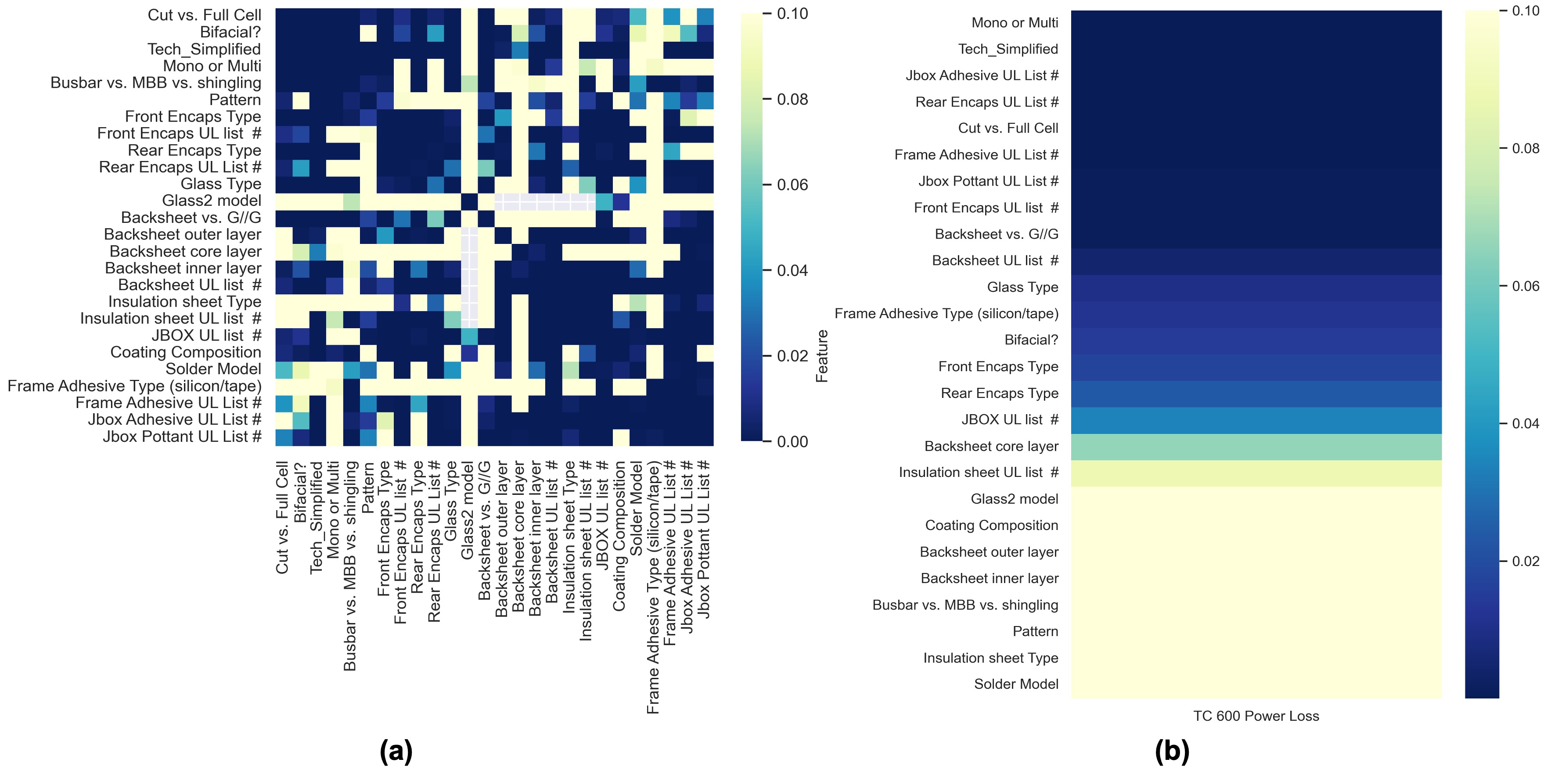}
\caption{Statistical association matrix among categorical features and the target variable. (a) The p-value of Chi-square test among categorical features. (b) The p-value of ANOVA test between features and the power loss. For both statistical testing, a low p-value indicates a potential dependence.}
\label{fig:corr}
\end{figure}

Figure~\ref{fig:spear-corr} demonstrates the full correlation matrix of numerical features. Figure~\ref{fig:corr} illustrates the statistical association among categorical features and that between categorical features and the target numerical variable. We use Chi-square test for categorical features, and the p-value of the testing is shown in Figure~\ref{fig:corr}(a). For the dependence between the categorical and the numerical target variable, we use ANOVA test, and the p-value is shown in Figure~\ref{fig:corr}(b). We note that this method is only used as a reference to select features but does not necessarily indicate correlation due to many confounding variables in the data set. Also, we do not include all the features that show potential correlations. For example, some features are redundant like ``Cut vs. Full Cell'' which can be inferred from the height and width of the cell. It should be noted that the domain knowledge described in the main paper is the primary standard for constructing the feature matrix.

\subsection{Full Feature Distribution}
\begin{figure}[H]
\centering
\includegraphics[width=0.9\textwidth]{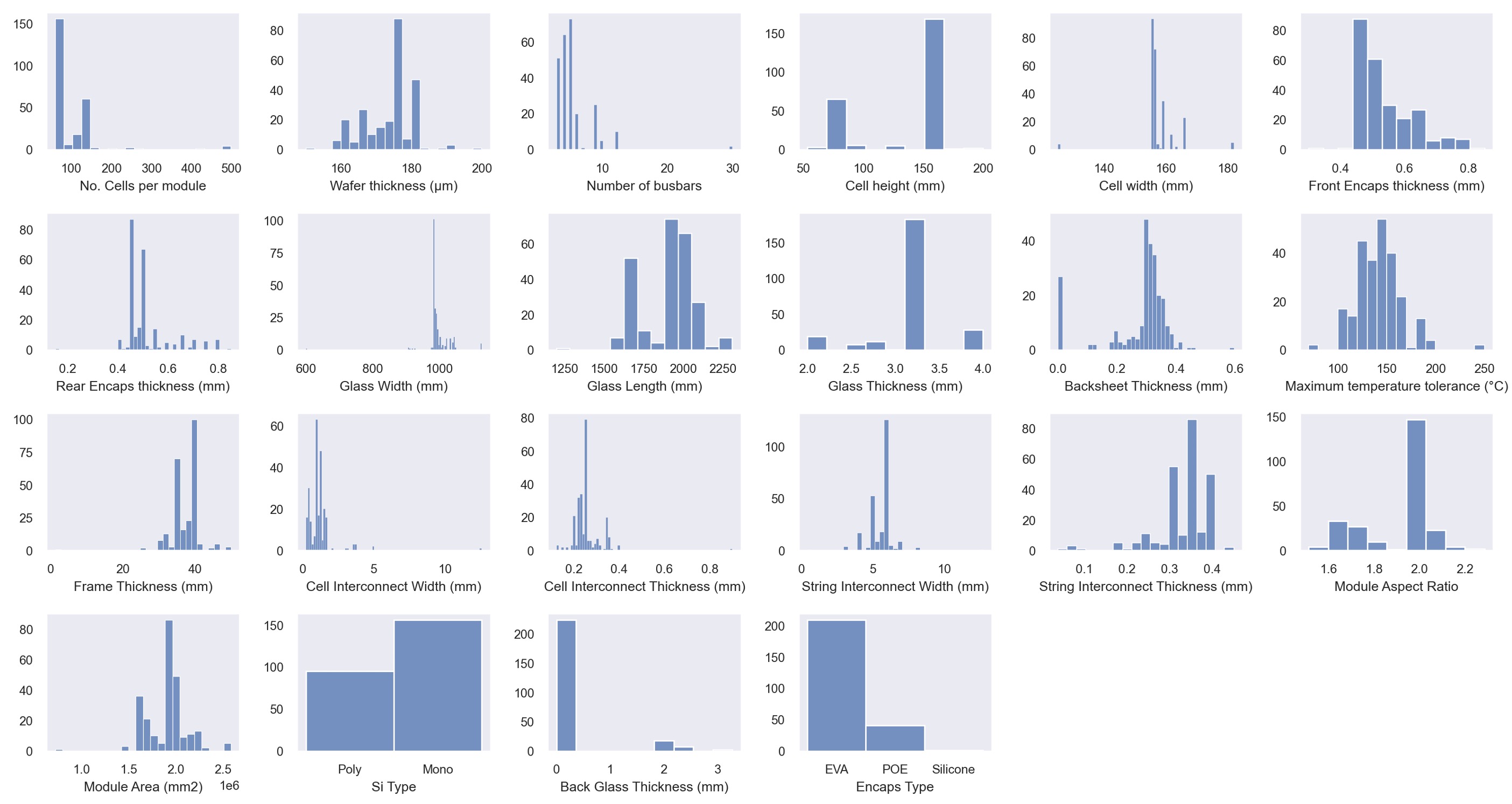}
\caption{Distribution of the 22 BOM features used to develop the ML models.}
\label{fig:feature_distribution}
\end{figure}

Figure~\ref{fig:feature_distribution} demonstrates the distribution of all the 22 features used to construct the feature matrix. The back glass thickness is the same as the front glass thickness for glass/glass modules and set as $0$ for glass/backsheet modules. Likewise, the backsheet thickness is the original value for glass/backsheet modules but set as $0$ for glass/glass modules.

\section{ML Modeling}
\subsection{L1 and L2 Regularization}
The $L1$ and $L2$ regularization terms follow the equations:

\begin{gather}
L1: \|\mathbf{w}\|_1=\sum_{i=1}^d\left|w_i\right|
\\
L2: \|\mathbf{w}\|_2^2=\sum_{i=1}^d w_i^2
\end{gather}
where $d$ is the number of features.
\subsection{Support Vector Regression}
The Support vector regression follows the equation:

\begin{equation}
\begin{gathered}
w, b = \operatorname{argmin}_{w, b, \zeta, \zeta^*} \frac{1}{2} \|w\|_2^2+C \sum_{i=1}^n\left(\zeta_i+\zeta_i^*\right) \\
\text { s.t. } y_i-w^T \phi\left(x_i\right)-b \leq \varepsilon+\zeta_i \\
w^T \phi\left(x_i\right)+b-y_i \leq \varepsilon+\zeta_i^* \\
\zeta_i, \zeta_i^* \geq 0, i=1, \ldots, n
\end{gathered}
\label{svr}
\end{equation}
where $w, b$ defines the hyperplane. Samples that are $\epsilon$ away from the hyperplane are penalized and $\zeta_i, \zeta_i^*$ control the penalization. $C$ is another hyperparameter that controls the regularization strength. $\phi\left(x_i\right)$ uses a kernel method that transforms the feature matrix to other dimensions. 

\subsection{Hyperparameter Tuning}
We perform hyperparameter tuning when training each model in the cross-validation process. The parameters we tuned and the optimal value for each parameter are listed in Table~\ref{tab:hyperparameters}. 
\begin{table}[H]
\centering
\caption{Hyperparameters used to develop ML models}
\begin{tabular}{|l|l|l|l|}
\hline
                         & Parameter           & Values for tuning                                                                                               & Optimal values \\ \hline
\multirow{2}{*}{Lasso}   & alpha               & {[}0.01, 0.1, 1, 10, 100, 1000{]}                                                                               & 0.1            \\ \cline{2-4} 
                         & selection           & {[}`cyclic', `random'{]}                                                                                        & `random'       \\ \hline
\multirow{2}{*}{Ridge}   & alpha               & {[}0.01, 0.1, 1, 10, 100, 1000{]}                                                                               & 1000           \\ \cline{2-4} 
                         & solver              & \begin{tabular}[c]{@{}l@{}}{[}`auto', `svd', `cholesky', `lsqr',\\  `sparse\_cg', `sag', `saga'{]}\end{tabular} & `svd'          \\ \hline
\multirow{4}{*}{SVR}     & kernel              & {[}`rbf'{]}                                                                                                     & `rbf'          \\ \cline{2-4} 
                         & C                   & np.logspace(-1, 3, 5)                                                                                           & 100            \\ \cline{2-4} 
                         & gamma               & np.logspace(-2, 2, 5)                                                                                           & 10            \\ \cline{2-4} 
                         & epsilon             & {[}0.1, 0.2, 0.3, 0.5{]}                                                                                        & 0.2            \\ \hline
\multirow{6}{*}{RF}      & n\_estimators       & \begin{tabular}[c]{@{}l@{}}{[}5, 10, 15{]}\end{tabular}                        & 15             \\ \cline{2-4} 
                         & max\_depth          & \begin{tabular}[c]{@{}l@{}}{[}1, 3, 5, 7, 9, 15{]}\end{tabular}                  & 15             \\ \cline{2-4} 
                         & min\_samples\_split & {[}2, 5, 15{]}                                                                                                  & 2              \\ \cline{2-4} 
                         & min\_samples\_leaf  & {[}1, 2, 4{]}                                                                                                   & 1              \\ \cline{2-4} 
                         & bootstrap           & {[}True, False{]}                                                                                               & True          \\ \cline{2-4} 
                         & max\_features       & {[}0.5, `sqrt', `log2'{]}                                                                                       & `sqrt'         \\ \hline
\multirow{4}{*}{XGBoost} & max\_depth          & \begin{tabular}[c]{@{}l@{}}{[}1, 3, 5, 7, 9, 15, 25, 35,\\  50, 60, 70, 80, 90{]}\end{tabular}                  & 7              \\ \cline{2-4} 
                         & n\_estimators       & \begin{tabular}[c]{@{}l@{}}{[}5, 10, 20, 30, 40, 50, 70, 80, 100,\\  200, 300, 400, 500, 1000{]}\end{tabular}   & 200           \\ \cline{2-4} 
                         & learning\_rate      & {[}0.01, 0.05, 0.1{]}                                                                                           & 0.05           \\ \cline{2-4} 
                         & gamma               & {[}0, 0.3, 0.5, 1, 5{]}                                                                                         & 5              \\ \hline
\end{tabular}
\label{tab:hyperparameters}
\end{table}

\subsection{Influence of Data Set Standardization}
\begin{figure}[H]
\centering
\includegraphics[width=0.9\textwidth]{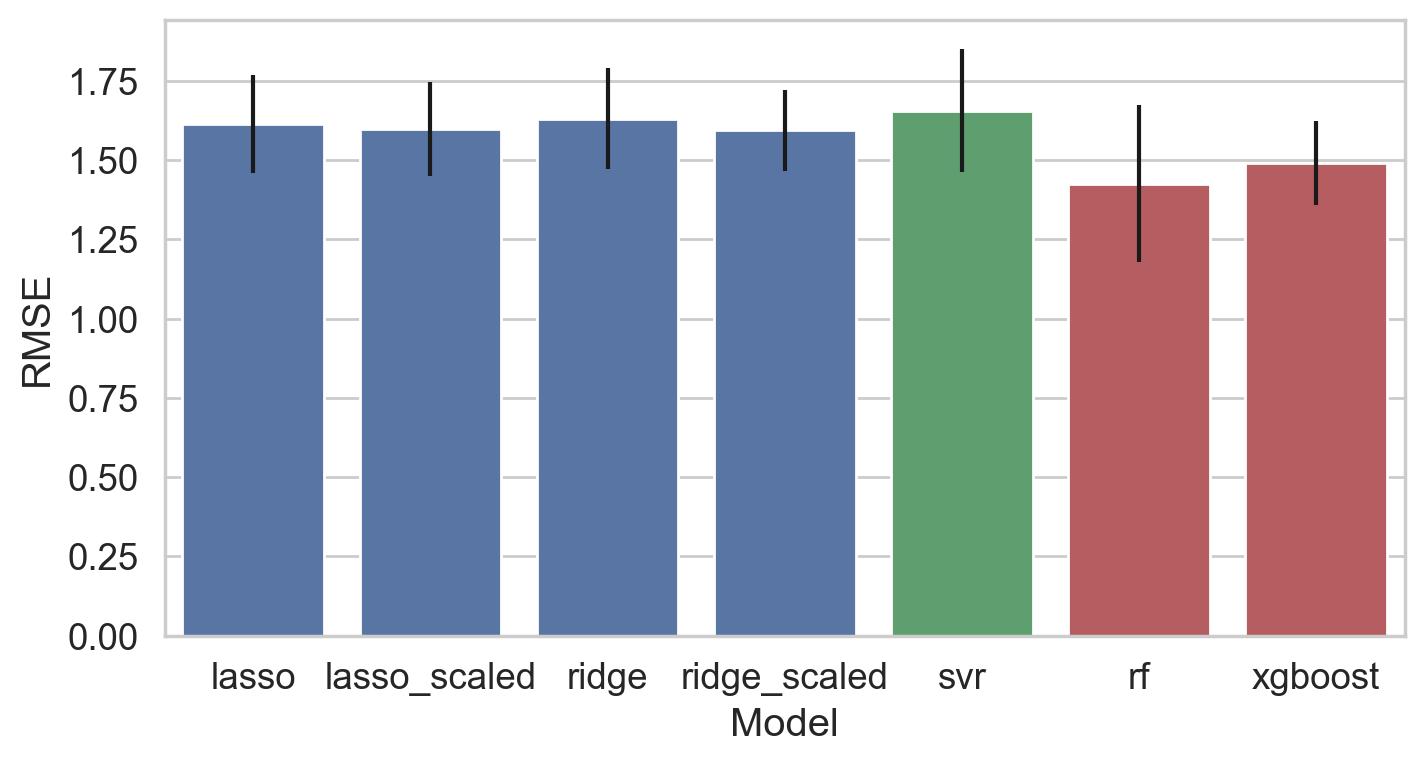}
\caption{Comparison of cross-validation RMSE scores among models. The data set is standardized for ``lasso\_scaled'' and ``ridge\_scaled'' models.}
\label{fig:scaled_ml}
\end{figure}

To counteract the influence of large differences between feature ranges on the linear models, we also train the linear models on the standardized data set. The training set is first standardized using ``StandardScaler'' function in Scikit-learn and then the testing set is standardized using the same ``StandardScaler'' parameters fitted on the training set. The comparison of the models using the standardized data set with other models is shown in Figure~\ref{fig:scaled_ml}. It can be seen that the standardization process doesn't significantly vary the linear models' performance and tree-based models are still the optimal models.

\subsection{Training and Testing Scores}
\begin{figure}[H]
\centering
\includegraphics[width=0.9\textwidth]{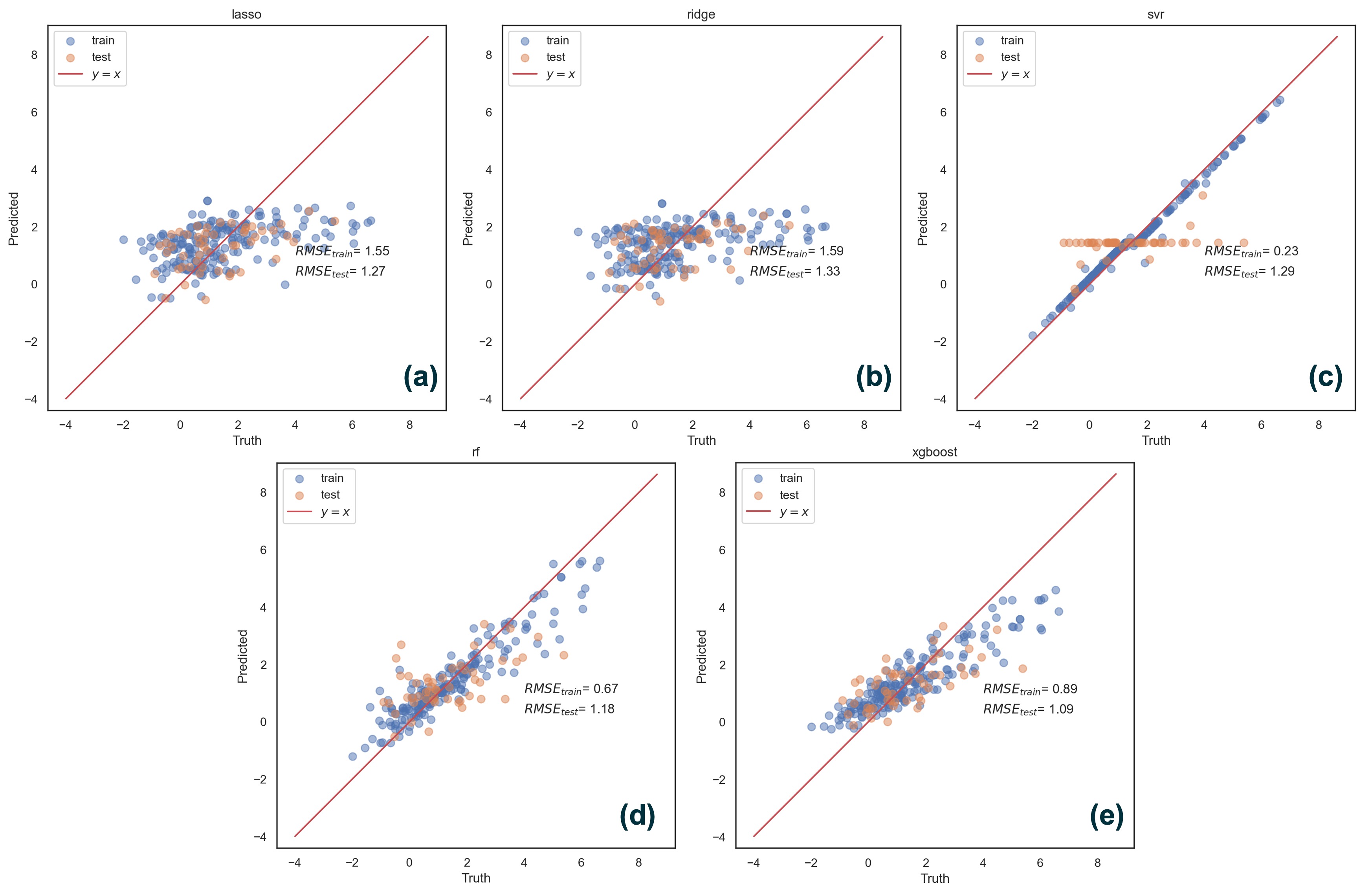}
\caption{RMSE of each model evaluated on both the training set and testing set. The deviation of the scatter points in linear and SVR models indicates a poor performance, although their RMSE scores are not significantly higher than tree-based models.}
\label{fig:test}
\end{figure}

The performance of all the models is shown in Figure~\ref{fig:test}. Both tree-based models (RF and XGBoost) illustrate good generalization with testing points distributed closely along $y=x$ line. However the points from other models deviate from this line. Especially, SVR shows nice performance during training but presents poor prediction power on the testing set by predicting the same value for different modules. This signifies the overfitting of SVR model. It should be noted that although the value of RMSE between tree-based models and other models is similar, the scatter plots shown in Figure~\ref{fig:test} probe the different prediction powers of each model. We note that the selection of optimal models is based on the cross-validation process rather than using the testing score. The testing scores of other models apart from RF shown here are only to demonstrate the characteristics of the scatter plots for overfitting and underfitting conditions.

\section{Post-hoc Statistical Analysis}
\subsection{Formula for T-test}
T-test is defined as the following equation:
\begin{equation}
t=\frac{\bar{a}-\bar{b}}{\sqrt{\frac{s_a^2}{n_a}+\frac{s_b^2}{n_b}}}
\end{equation}
where $\bar{a}, \bar{b}$ are the sample means, and $n_a, n_b$ are the sample sizes, and $s_a^2, s_b^2$ are the sample variances. 

\subsection{Clustering and Feature Dependence}

\begin{figure}[H]
\centering
\includegraphics[width=0.8\textwidth]{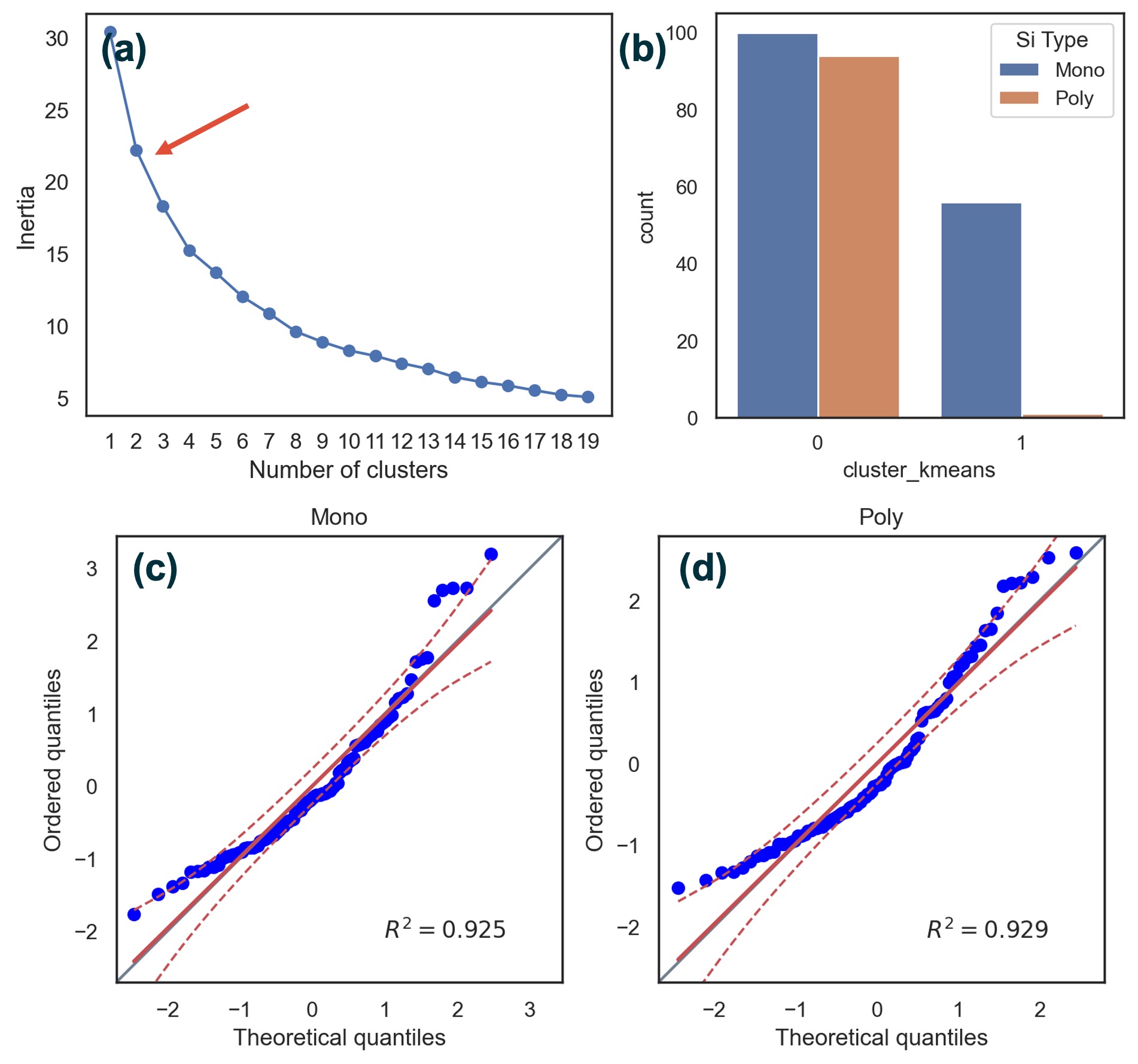}
\caption{Statistical testing to verify the impact of the Si cell type. (a) The elbow method used to determine the number of clusters in KMeans clustering. ``number of clusters=2'' (pointed by the red arrow) is used in this study. (b) The distribution of cells in the two clusters after regrouping. Cluster 0 is selected due to larger number of cells. (c)(d) The QQ-plot used to check the normality of the power loss distribution, which is required to satisfy the assumption of t-test. }
\label{fig:cluster_si}
\end{figure}

Figure~\ref{fig:cluster_si} shows the clustering and normality check to test the Si type. We first use the elbow method to determine the number of clusters by computing the inertia for different numbers of clusters. Inertia measures the distance between each point and the centroid of its cluster. A good clustering model should have both low inertia and a low number of clusters, which is known as the elbow position. It should be noted that the selection of the elbow is subjective. In our study, we select the elbow as long as the features apart from the target feature such as the Si type tested in the main paper can be controlled in the regrouped subset. Figure~\ref{fig:cluster_si}(a) illustrates that we selected ``number of clusters=2'' as the elbow. Figure~\ref{fig:cluster_si}(b) shows the distribution of the number of cells in each cluster when we reorganize the original data set into two subsets using the clustering method. Here we test our null hypothesis on cluster 0 since it contains enough poly-c and mono-c cells. The Q-Q plots in Figure~\ref{fig:cluster_si}(c)(d) show that the high correlation coefficient ($>0.9$) between theoretical quantiles and ordered quantiles for both mono-c and poly-c Si cells indicates that the power loss of these two groups of cells is normally distributed. 

\begin{figure}[H]
\centering
\includegraphics[width=0.9\textwidth]{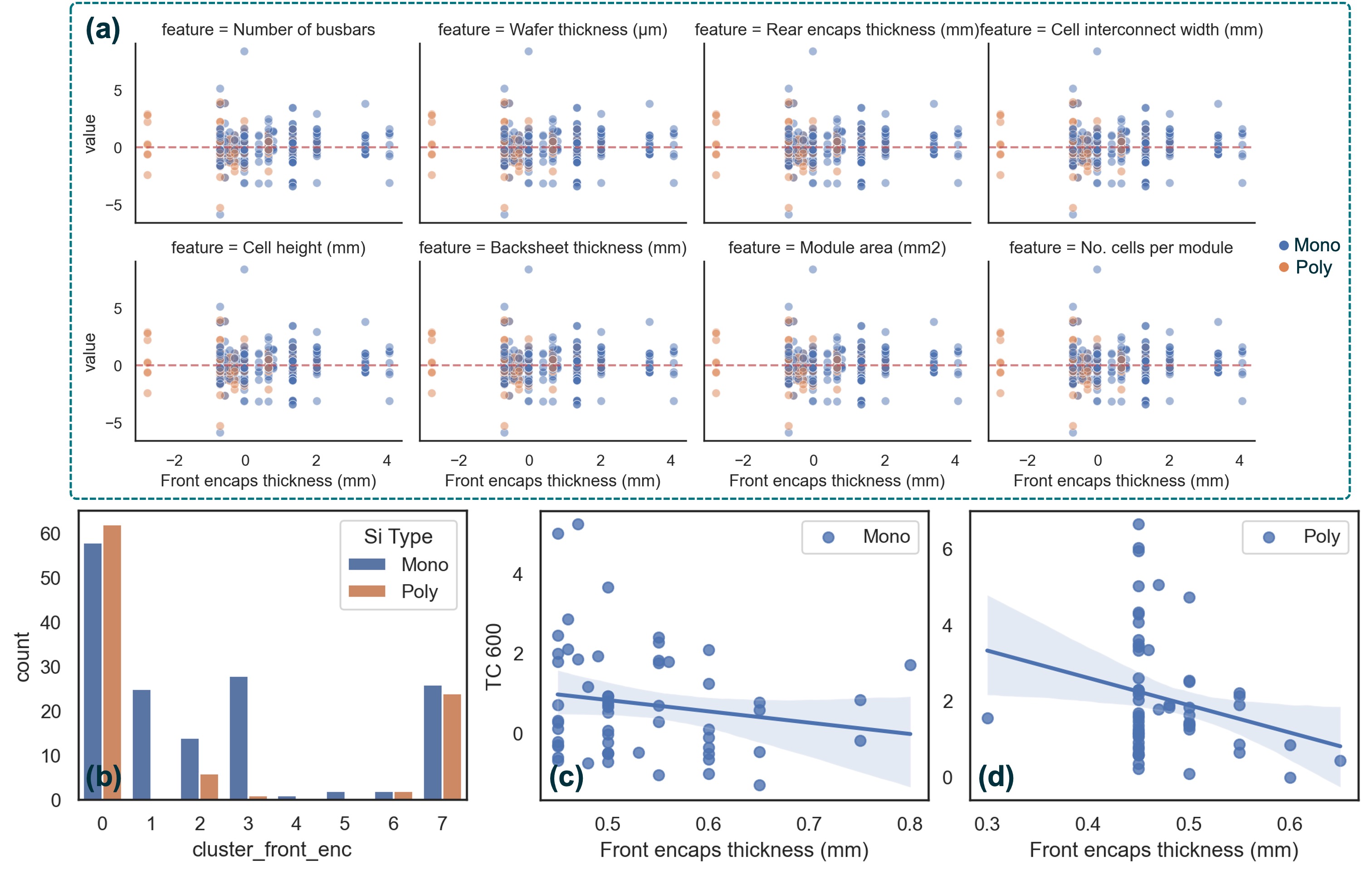}
\caption{Statistical testing to verify the impact of the front encapsulant thickness. Sample data was first extracted from the raw data set using the clustering method to guarantee that features other than front encapsulant thickness are controlled. We select the number of clusters to be 8 and use cluster 0 for testing. (a) The normalized values of other controlled attributes in the selected cluster 0. Those variables are not well controlled because it is more difficult to regroup the data set based on numerical feature (\emph{i.e.,} front encapsulant thickness). (b) The distribution of number of cells in cluster 0, compared with other clusters. (c)(d) The regression lines for the the front encapsulant impact. Poly-c and mono-c Si cells are tested separately. P-val for the regression in (c) is 0.261 and for (d) is 0.270.}
\label{fig:cluster_front_encap}
\end{figure}

\begin{figure}[H]
\centering
\includegraphics[width=0.9\textwidth]{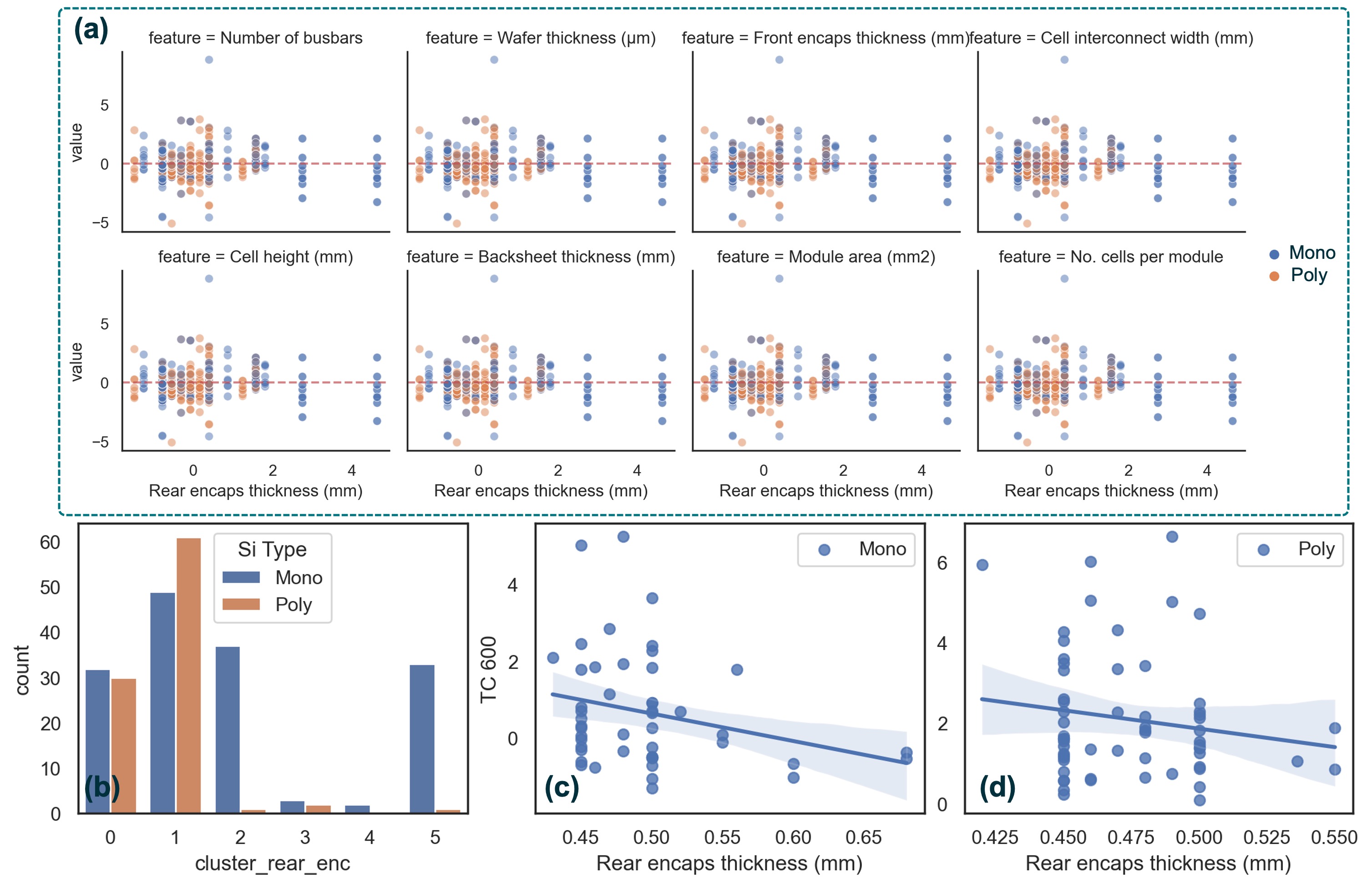}
\caption{Statistical testing to verify the impact of the rear encapsulant thickness. Sample data was first extracted from the raw data set using the clustering method to guarantee
that features other than rear encapsulant thickness are controlled. We select the number of clusters to be 6 and use cluster 1 for testing. (a) The normalized values of other controlled attributes in cluster 1. Those variables are not well controlled because it is more difficult to regroup the data set based on numerical feature (\emph{i.e.,} rear encapsulant thickness). (b) The distribution of number of cells in cluster 1, compared with other clusters. (c)(d) The regression lines for the the rear encapsulant impact. Poly-c and mono-c Si cells are tested separately. P-val for the regression in (c) is 0.087 and for (d) is 0.489.}
\label{fig:cluster_rear_encap}
\end{figure}

\begin{figure}[H]
\centering
\includegraphics[width=0.9\textwidth]{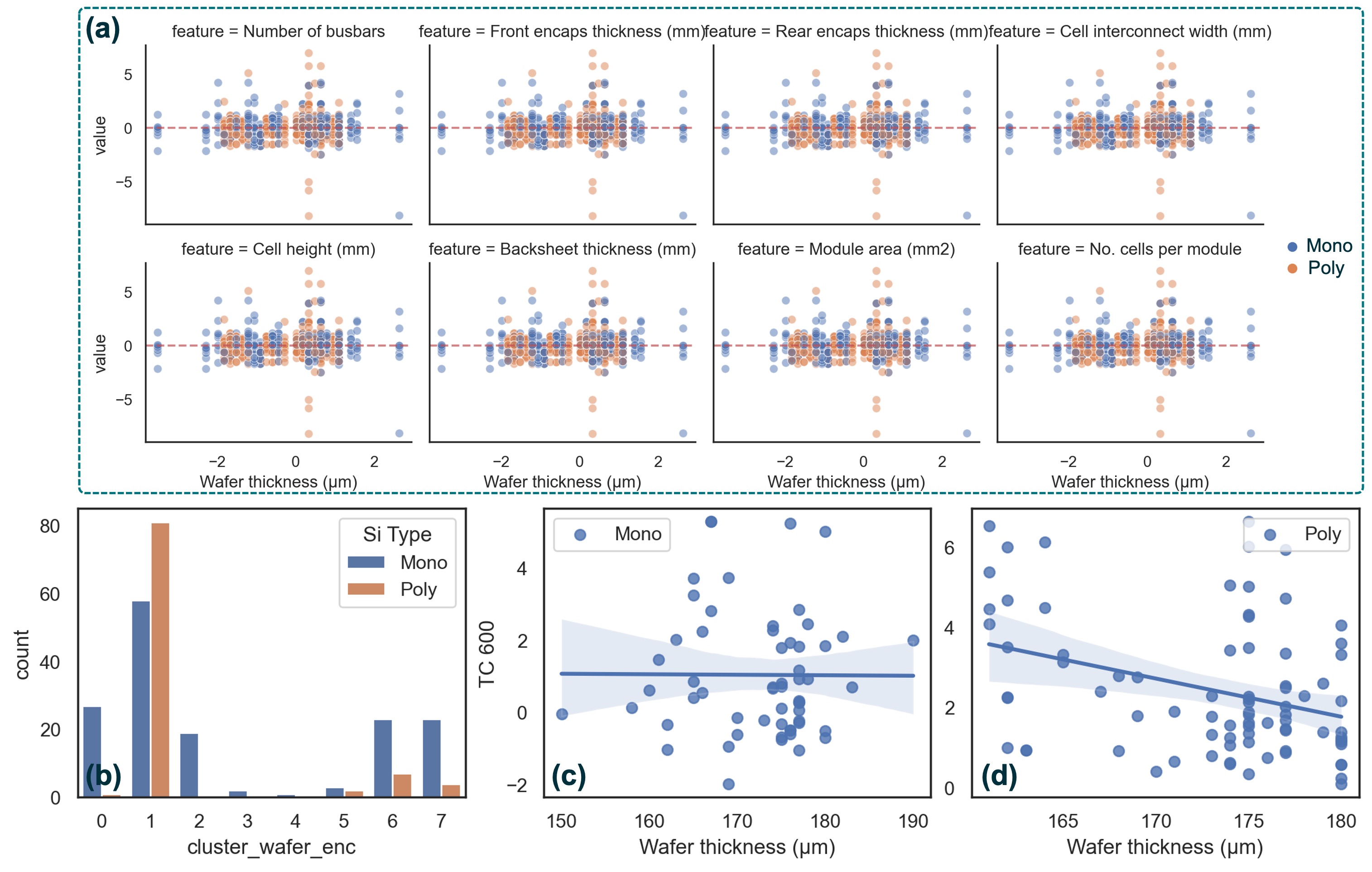}
\caption{Statistical testing to verify the impact of the wafer thickness. Sample data was first extracted from the raw data set using the clustering method to guarantee
that features other than wafer thickness are controlled.We select the number of clusters to be 8 and use cluster 1 for testing. (a) The normalized values of other controlled attributes in cluster 1. Those variables are not well controlled because it is more difficult to regroup the data set based on numerical feature (\emph{i.e.,} wafer thickness). (b) The distribution of number of cells in cluster 1, compared with other clusters. (c)(d) The regression lines for the the wafer thickness impact. Poly-c and mono-c Si cells are tested separately. P-val for the regression in (c) is 0.937 and for (d) is 0.015.}
\label{fig:cluster_wafer}
\end{figure}

\begin{figure}[H]
\centering
\includegraphics[width=0.9\textwidth]{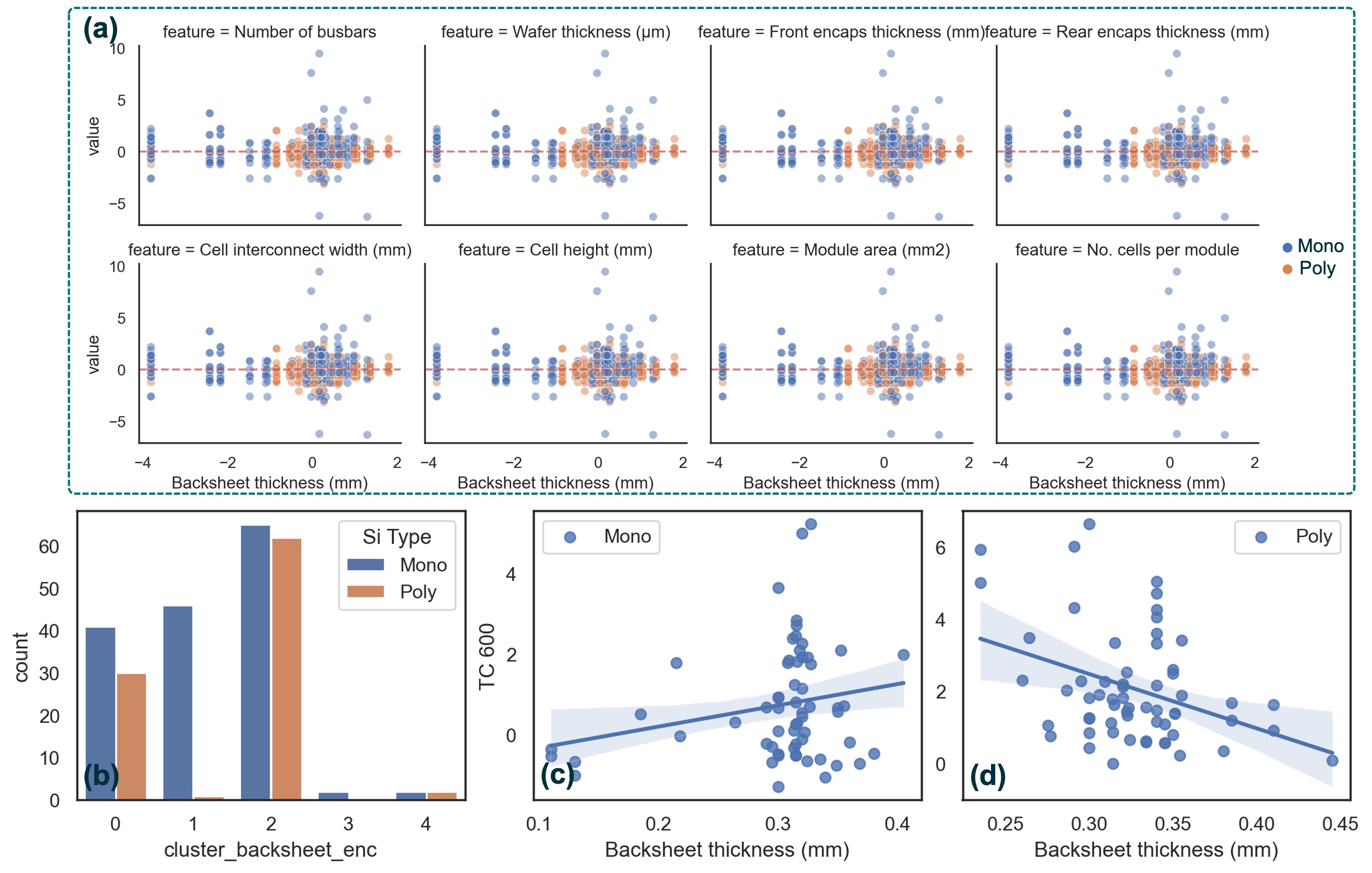}
\caption{Statistical testing to verify the impact of the backsheet thickness. Sample data was first extracted from the raw data set using the clustering method to guarantee that features other than backsheet thickness are controlled. We select the number of clusters to be 5 and use cluster 2 for testing.  (a) The normalized values of other controlled attributes in cluster 2. Those variables are not well controlled because it is more difficult to regroup the data set based on numerical feature (\emph{i.e.,} backsheet thickness). (b) The distribution of number of cells in cluster 2, compared with other clusters. (c)(d) The regression lines for the the backsheet thickness impact. Poly-c and mono-c Si cells are tested separately. P-val for the regression in (c) is 0.176 and for (d) is 0.042.}
\label{fig:cluster_backsheet}
\end{figure}

In the main paper, the clustering method is effectively applied to categorical data (\emph{i.e.,} the Si type) to validate the SHAP interpretation that poly-c Si modules have higher power loss than mono-c Si modules. However, this grouping method cannot achieve the same level of controlled grouping to test numerical variables. Figures~\ref{fig:cluster_front_encap}, \ref{fig:cluster_rear_encap}, \ref{fig:cluster_wafer} and \ref{fig:cluster_backsheet} demonstrate statistical testing of some features, including front encapsulant thickness, rear encapsulant thickness, wafer thickness, and backsheet thickness. The feature values in those plots are normalized for better visualization. It can be seen that features are not well controlled in Figure~\ref{fig:cluster_front_encap}(a), \ref{fig:cluster_rear_encap}(a), \ref{fig:cluster_wafer}(a) and \ref{fig:cluster_backsheet}(a) since the normalized values for these features are not consistent as the primary variable changes. Consequently, these results are not robust enough to support the SHAP interpretation and, therefore, are not included in the main paper's analysis. Despite this limitation, the regression results of Figure~\ref{fig:cluster_front_encap}(c)(d), \ref{fig:cluster_rear_encap}(c)(d), \ref{fig:cluster_wafer}(d) and \ref{fig:cluster_backsheet}(c) still reveal trends that agree with the SHAP interpretation. Although the clustering of numerical features does not provide a perfect framework for validation, these trends here still offer indirect support to the SHAP results.

\bibliographystyle{IEEEtran}
\bibliography{refs.bib}